\documentclass[times,twocolumn,final]{elsarticle}

\usepackage{medima}
\usepackage{framed,multirow}

\usepackage{amssymb}
\usepackage{latexsym}

\usepackage{url}
\usepackage{xcolor}
\usepackage{makecell}
\usepackage{tabu, booktabs}
\usepackage{hyperref}
\usepackage{lineno}
\definecolor{newcolor}{rgb}{.8,.349,.1}

\journal{Medical Image Analysis}

\begin{document}

\verso{Luo \textit{et~al.}}

\begin{frontmatter}

\title{WORD: A large scale dataset, benchmark and clinical applicable study for abdominal organ segmentation from CT image}

\author[1,4]{Xiangde Luo}
\author[2,5]{Wenjun Liao}
\author[3]{Jianghong Xiao\corref{cores}}\ead{xiaojh@scu.edu.cn}
\author[6]{Jieneng Chen}
\author[7]{Tao Song}
\author[4]{Xiaofan Zhang}
\author[8]{Kang Li}
\author[9]{Dimitris~N.~Metaxas}
\author[1,4]{Guotai Wang\corref{cores}}\ead{guotai.wang@uestc.edu.cn}
\author[1,4]{Shaoting Zhang\corref{cores}}\ead{Rutgers.shaoting@gmail.com}\cortext[cores]{Corresponding authors}
\address[1]{School of Mechanical and Electrical Engineering, University of Electronic Science and Technology of China, Chengdu, China}
\address[2]{Department of Radiation Oncology, Sichuan Cancer Hospital and Institute, Sichuan Cancer Center, Chengdu, China}
\address[3]{Department of Radiation Oncology, Cancer Center West China Hospital, Sichuan University, Chengdu, China}
\address[4]{Shanghai Artificial Intelligence Laboratory, Shanghai, China}
\address[5]{School of Medicine, University of Electronic Science and Technology of China, Chengdu, China}
\address[6]{Department of Computer Science, Johns Hopkins University, Baltimore, USA}
\address[7]{SenseTime Research, Shanghai, China}
\address[8]{West China Hospital-SenseTime Joint Lab, West China Biomedical Big Data Center, Sichuan University, Chengdu, China}
\address[9]{Department of Computer Science, Rutgers University, Piscataway, NJ USA}





\begin{abstract}
Whole abdominal organ segmentation is important in diagnosing abdomen lesions, radiotherapy, and follow-up. However, oncologists' delineating all abdominal organs from 3D volumes is time-consuming and very expensive. Deep learning-based medical image segmentation has shown the potential to reduce manual delineation efforts, but it still requires a large-scale fine annotated dataset for training, and there is a lack of large-scale datasets covering the whole abdomen region with accurate and detailed annotations for the whole abdominal organ segmentation. In this work, we establish a new large-scale \textit{W}hole abdominal \textit{OR}gan \textit{D}ataset (\textit{WORD}) for algorithm research and clinical application development. This dataset contains 150 abdominal CT volumes (30495 slices). Each volume has 16 organs with fine pixel-level annotations and scribble-based sparse annotations, which may be the largest dataset with whole abdominal organ annotation. Several state-of-the-art segmentation methods are evaluated on this dataset. And we also invited three experienced oncologists to revise the model predictions to measure the gap between the deep learning method and oncologists. Afterwards, we investigate the inference-efficient learning on the WORD, as the high-resolution image requires large GPU memory and a long inference time in the test stage. We further evaluate the scribble-based annotation-efficient learning on this dataset, as the pixel-wise manual annotation is time-consuming and expensive. The work provided a new benchmark for the abdominal multi-organ segmentation task, and these experiments can serve as the baseline for future research and clinical application development.
\end{abstract}


\end{frontmatter}

\begin{table*}[t]
    \centering
    \scalebox{0.83}{\begin{tabu}{c|c|c|c|c|c|c}
    \hline
    Dataset& Year& Organs& Scans& NSD& AFS& WAR\\
    \hline
    BTCV~\citep{landman2017multi}&2015& \makecell[cc]{Spleen, Kidney (L), Kidney (R), Gallbladder, Esophagus, \\ Liver, Stomach, Aorta, Inferior vena cava, Portal vein and splenic vein, \\ Pancreas, Adrenal gland(L), Adrenal gland(R), Duodenum}& 50& \textcolor{red}{50} & $\surd$& $\times$ \\
    \hline
    DenseVNet~\citep{gibson2018automatic}&2018& \makecell[cc]{Spleen, Kidney (R), Gallbladder, Esophagus, Liver, \\ Stomach, Splenic vein, Pancreas and Duodenum}& 90& $\times$& $\surd$& $\times$\\
    \hline 
    CT-ORG~\citep{rister2020ct}&2020& \makecell[cc]{Lung, Bones, Liver, Kidneys and Bladder}& 140& $\times$& $\times$& $\times$ \\
    \hline
    AbdomenCT-1K~\citep{ma2021abdomenct}&2021& \makecell[cc]{Spleen, Kidney, Liver and Pancreas}& 1112& \textcolor{red}{50}& $\times$& $\times$\\
    \hline
    WORD dataset (ours)&2022&\makecell[cc]{Liver, Spleen, Kidney(L), Kidney(R), Stomach, Gallbladder, Esophagus,\\ Pancreas, Duodenum, \textbf{Colon}, \textbf{Intestine}, Adrenal(L), Adrenal(R), \textbf{Rectum},\\ Bladder, Head of Femur(L) and Head of Femur(R)}&170& \textcolor{red}{150} & \textbf{$\surd$}& \textbf{$\surd$}\\
    \hline
    \end{tabu}}
    \caption{Summary of several publicly available abdominal CT datasets. NSD: New Source Data \textcolor{red}{(total scans)}; AFS: Annotate From Scratch. WAR: with the Whole Abdominal Region. To the best of our knowledge, WORD dataset is the first whole abdominal organ dataset. \textcolor{red}{We corrected the results or description by the red text.}}
    \label{tab:datasets_summary}
\end{table*}

\section{Introduction}
Abdominal organ segmentation is a fundamental and essential task in abdominal disease diagnosis, cancer treatment, and radiotherapy planning~\citep{tang2019clinically}. As accurate segmentation results can provide pieces of valuable information for the clinical diagnosis and follow-ups, like organ size, location, boundary state, the spatial relationship of multiple organs, etc. In addition, organ segmentation plays a critical role in clinical treatment, especially in radiation therapy-based cancer and oncology treatments~\citep{chen2021deep}. Accurate segmentation of organs at risk can alleviate potential effects on healthy organs near cancer regions. However, in clinical practice, organ segmentation is usually manually performed by radiation oncologists or radiologists. It is time-consuming and error-prone, requiring annotators to delineate and check slice-by-slice and may take several hours per case. In addition, due to the different imaging protocols/quality, and anatomical structures, fast delineation of many organs is also a challenging task for junior oncologists~\citep{guo2020organ}.
\begin{figure}[t]
    \centering
    \includegraphics[width=0.48\textwidth]{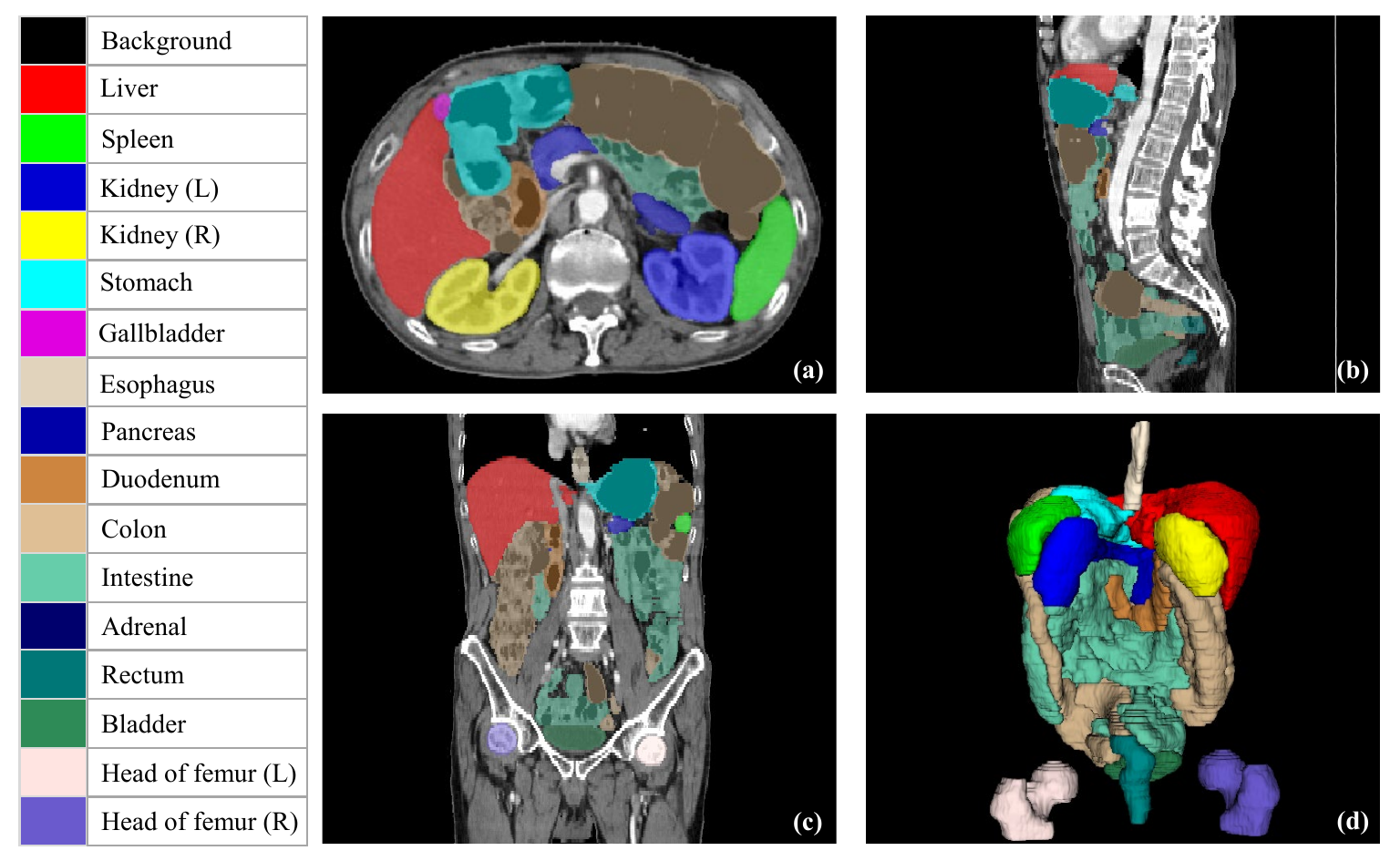}
    \caption{{An example of 16 annotated abdominal organs in a CT scan.} The left table lists the annotated organs' categories. (a), (b), (c) denote the visualization in axial, coronal, and sagittal views, respectively. (d) represents the 3D rendering results of annotated abdomen organs.}
    \label{fig:show_data}
\end{figure}

\par {Recently, many deep learning-based methods have been proposed to accurately and quickly segment organs from abdominal CT volumes~\citep{chen2021deep,ma2021abdomenct,wang2019abdominal}. However, these methods were evaluated on small or in-house datasets or just segmented very few organs. In addition, previous works have also shown that some abdominal organ segmentation has achieved very promising results, such as liver, spleen, and kidney~\citep{ma2021abdomenct}. But there are still some abdominal organ segmentation tasks that are unsolved and challenging, especially for small and complex organs~\citep{ma2021abdomenct,chen2021deep}. The main reason caused these problems may be lacking a publicly available large-scale real clinical dataset with accurate whole abdominal organ annotation for research. So, developing high-quality and large-scale datasets and building benchmarks for the whole abdominal organ segmentation task is vital to boost these unsolved organ segmentation studies~\citep{ma2021abdomenct,chen2021deep}.}

\par In this work, our goal is to collect a large-scale real clinical abdomen dataset (WORD) with careful annotations. All scans in our dataset are manually segmented in great detail, covering 16 organs in the abdominal region. Due to privacy and ethical protection, collecting real clinical data is challenging and time-consuming. In addition, annotating a large-scale 3D medical image segmentation dataset is very expensive and labour-intensive, as it requires domain knowledge and clinical experience. Recently, some researchers reused previous datasets by providing annotations with pre-trained models or semi-automatic methods~\citep{ma2021abdomenct,rister2020ct}, which may affect the annotator's decision, especial regarding low-contrast boundary regions. In contrast, WORD dataset was collected from a radiation therapy center and annotated by one senior oncologist (with 7 years of experience) and then checked, discussed and refined by an experts (more than 20 years of experience). All of images were scanned before the radiotherapy treatment, without any appearance enhancement, with a similar scan location and with a similar image spacing, etc. Figure~\ref{fig:show_data} shows an example from WORD.

\par Moreover, we investigate current methods on the WORD dataset, including fully supervised segmentation and annotation-efficient methods. Specifically, we first evaluate several state-of-the-art medical segmentation methods on the WORD, like Convolutional Neural Network-based methods nnUNet~\citep{isensee2021nnu}, Attention UNet~\citep{oktay2018attention}, DeepLabV3+~\citep{chen2018encoder}, UNet++~\citep{zhou2019unet++} and ResUNet~\citep{diakogiannis2020resunet}, and transformer-based approaches like CoTr~\citep{xie2021cotr} and UNETR~\citep{hatamizadeh2022unetr}. After that, we investigate generalization ability of a pre-trained model on the BTCV~\citep{landman2017multi} and TCIA~\citep{roth2015deeporgan}. Due to previous datasets only have annotations of few organs, we further annotate an open dataset for generalization ability evaluation, where 20 cases with the whole abdominal region were selected and annotated manually from the LiTS~\citep{bilic2019liver} dataset. Afterwards, we do the user study on this dataset to measure the gap between deep learning models and three oncologists. Considering these CT images have very high resolution, we investigate inference-efficient learning to reduce the memory and time cost and accelerate the inference procedure. Finally, we introduce a weakly supervised abdominal organ segmentation method with scribble annotations, which is desirable to reduce the annotation cost in the future. These attempts can be used as a new abdominal organ segmentation benchmark for further research. In summary, our contribution is two-fold:
\begin{itemize}
    \item[1)] We build a new clinical whole abdominal organ segmentation dataset (150 CT scans) and has more categories (16 organs) and high-quality annotations than previous works~\citep{landman2017multi,gibson2018automatic,rister2020ct,ma2021abdomenct}. In addition, we further annotate 20 cases from LiTS~\citep{bilic2019liver} for networks' generalization evaluation~\footnote{\url{https://github.com/HiLab-git/WORD}}.
    \item[2)] We establish a new abdominal organ segmentation benchmark by (1) evaluating the existing fully supervised segmentation methods, (2) measuring the gap between deep learning models and oncologists, (3) investigating the pre-trained model generalization ability on open datasets, (4) investigating the inference-efficient learning for the high-resolution abdominal CT image segmentation, (5) introducing scribble-based weakly supervised methods to reduce the labeling cost.
\end{itemize}
\section{Related work}
\begin{figure*}
    \centering
    \includegraphics[width=0.99\textwidth]{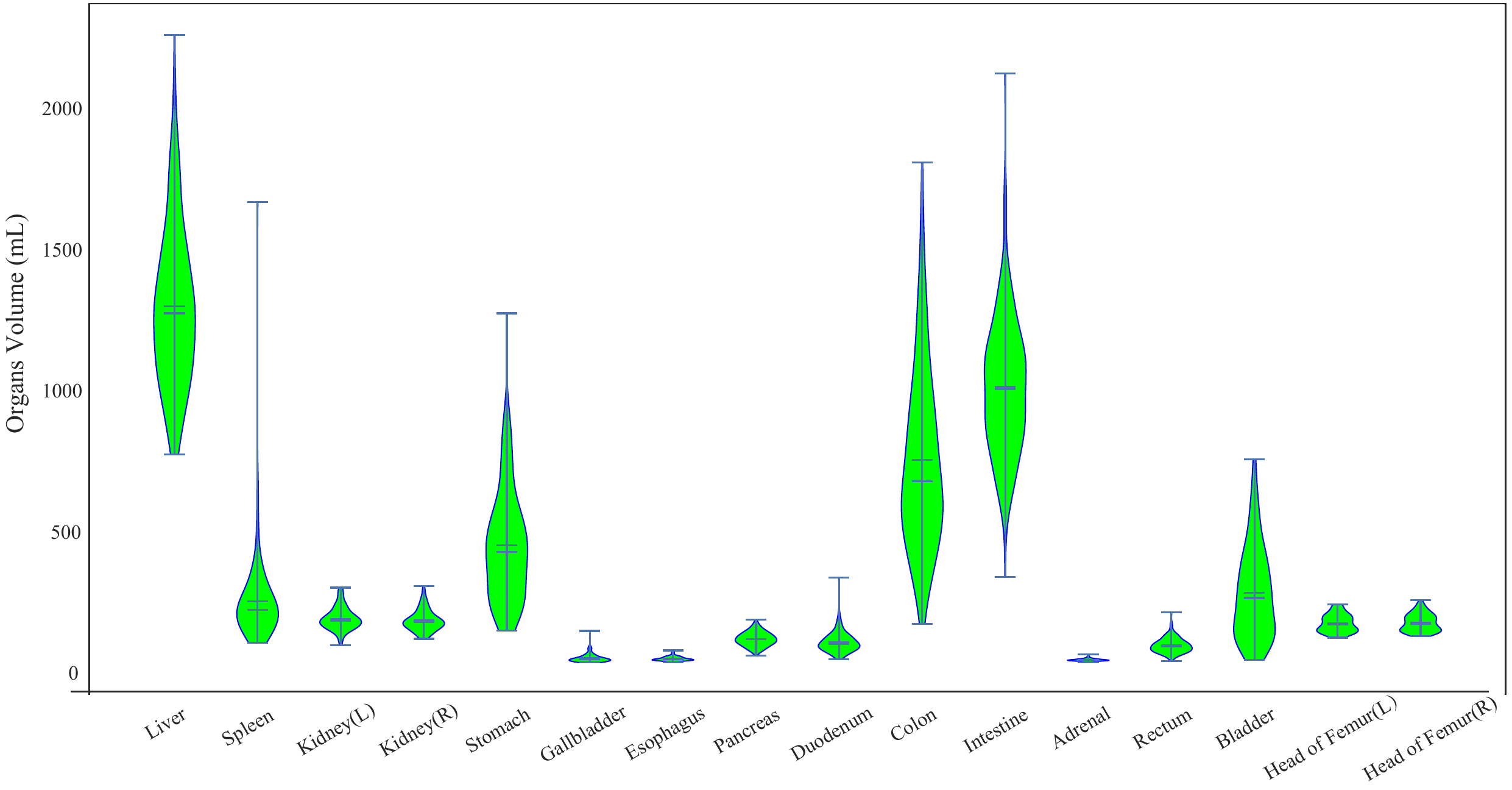}
    \caption{Volume distribution of 16 organs in WORD. 
    }
    \label{fig:data_dist}
\end{figure*}
\subsection{abdominal organ segmentation datasets}
Since clinical CT images of the whole abdominal region are very private and challenging to collect and annotate, few datasets with carefully annotated whole abdominal organs are publicly available. We summarize these publicly available abdominal CT datasets in Table~\ref{tab:datasets_summary}. We consider the datasets with four or more annotated organs in this work. The BTCV (Beyond The Cranial Vault)~\citep{landman2017multi} consists of 50 CT volumes, with 30 and 20 volumes used for training and testing, respectively. In the BTCV dataset, 13 organs are annotated manually, including the aorta, liver, spleen, right kidney, left kidney, stomach, pancreas, gallbladder, esophagus, inferior vena cava, portal vein and
splenic vein, right adrenal gland, and left adrenal gland. The DenseVNet~\citep{gibson2018automatic} has 90 CT scans, where 47 scans come from the BTCV dataset~\citep{landman2017multi}, and the other 43 cases come from TCIA data~\citep{roth2015deeporgan} each with annotations of eight organs. The CT-ORG~\citep{rister2020ct} is an open dataset that contains 140 CT images, and five organs are annotated. Most of these images come from a challenge training set~\citep{bilic2019liver}. The AbdomenCT-1K dataset~\citep{ma2021abdomenct} extend five public singular organ segmentation datasets to four classes (with 1062 volumes) and a small clinical dataset (with 50 volumes coming from 20 patients). This dataset contains four organ annotations: liver, kidney, spleen, and pancreas. BTCV, DenseNet, and CT-ORG are limited by the small scale or few annotated classes to boost this topic research. Although AbdomenCT-1K is huge, the annotated organs are also too few to evaluate the efficiency of the whole abdominal segmentation task. Unlike these existing datasets, our dataset comes from a new medical center with a large scale and more annotated organs, such as the colon, intestine, rectum, etc. We believe WORD dataset is one of the most comprehensive datasets for medical image segmentation.
\subsection{abdominal organ segmentation methods}
Recently, deep learning-based methods have been widely used in abdominal organ segmentation tasks, especially the UNet-based deep networks~\citep{ronneberger2015u}. The main challenge in this task lies in complex anatomical structures, the unclear boundary of soft tissues, high resolution of images, and extremely unbalanced sizes among large and small organs, etc. Many works have attempted to handle these challenges.~\citet{gibson2018automatic} proposed a DenseVNet to segment 8 organs from CT volumes, which enables high-resolution activation maps through memory-efficient dropout and feature reuse.~\citet{wang2019abdominal} presented a novel framework for abdominal multi-organ segmentation using organ-attention networks with reverse connections and evaluated it on an in-house dataset.~\citet{liang2021incorporating} combined the inter-and intra-patient deformation data augmentation with multi-scale Attention-UNet~\citep{schlemper2019attention} for accurate abdominal multi-organ segmentation.~\citet{tang2021high} proposed a batch-based method plus random shifting strategy to boost the performance of multi-organ segmentation from high-resolution abdomen CT volumes. More recently, transformer-based methods~\citep{cao2021swin,chen2021transunet} are used to explicitly model the long-range dependence to capture the relationship of multi-organ for accurate segmentation. 
\par Although the above methods have achieved promising results, they are also limited by requiring large scale carefully annotated dataset. To reduce annotation cost,~\citet{zhou2019semi} proposed a co-training-based semi-supervised method for abdominal multi-organ segmentation, which reduces almost half of annotation cost. Furthermore,~\citet{zhou2019prior} proposed a prior-aware neural network that incorporates anatomical priors on abdominal organ sizes to train models from several partially-labelled datasets. This work first investigates more annotation-efficient abdominal multi-organ segmentation methods with sparse annotations (scribbles). In addition, we investigate inference-efficient learning for the segmentation of high-resolution abdominal CT images to reduce the memory and time cost in the test stage. 

\section{WORD: Fully Annotated Clinical Whole Abdominal Organ Dataset}

\begin{table}
    \centering
    \scalebox{0.87}{\begin{tabular}{l|c|c|c}
    \hline
    Characteristics&Train (n=100) & Validation (n=20) & Test (n=30)\\
    \hline
    Age (median) & 47 (28-75) & 52 (32-78) & 49 (26-72) \\
    \hline
    Male& 63 & 12 & 13 \\
    Female& 37 & 8 & 17\\
    \hline
    Prostatic cancer&28&7& 10\\
    Cervical cancer&29&6&5\\
    Rectal cancer&26&3&8\\ 
    Others&17&4&7\\ 
    \hline
    \end{tabular}}
    \caption{Clinical characteristics of WORD. Others include some metastatic tumours, such as bone metastasis and soft tissue metastasis.}
    \label{tab:char_label}
\end{table}
\begin{table*}
\centering
\normalsize
\scalebox{0.50}{\begin{tabular}{l|c|c|c|c|c|c|c|c|c|c|c|c|c|c|c|c}
\hline
Organs & Liver& Spleen ~& ~ Kidney (L) ~& ~ Kidney (R)& ~ Stomach& ~ Gallbladder& ~Esophagus& ~ Pancreas& ~ Duodenum& Colon& Intestine& Adrenal& Rectum& Bladder& Head of Femur (L)& Head of Femur (R)\\
\hline
Senior&1.73$\pm$0.03 &
1.34$\pm$0.12 &
1.28$\pm$0.09 &
1.36$\pm$0.11 &
3.95$\pm$0.52 &
6.86$\pm$1.29 &
7.65$\pm$2.24 &
7.49$\pm$2.39 &
9.67$\pm$4.11 &
4.74$\pm$2.35 &
3.66$\pm$1.26 &
9.86$\pm$4.38 &
3.76$\pm$1.44 &
2.73$\pm$0.89 &
1.78$\pm$0.94 &
1.63$\pm$0.46\\
Expert&0.89$\pm$0.01 &
0.93$\pm$0.03 &
1.03$\pm$0.08 &
0.97.$\pm$0.07 &
1.48$\pm$0.35 &
2.97$\pm$0.94 &
3.32$\pm$1.25 &
2.89$\pm$0.91 &
3.33$\pm$1.23 &
1.27$\pm$0.23 &
1.48$\pm$0.69 &
3.75$\pm$1.73 &
1.46$\pm$0.74 &
1.35$\pm$0.39 &
0.96$\pm$0.13 &
0.87$\pm$0.09\\
\hline
\end{tabular}}
\caption{Quantitative analysis of the inconsistent ratio ($DSC~(\%)$) between the annotation of senior oncologist, the expert and their consensus annotations.}
\label{tab:revision_ratio}
\end{table*}

\subsection{Dataset summary}
The 150 CT scans in the WORD dataset were collected from 150 patients before the radiation therapy in a single center. All of them are scanned by a SIEMENS CT scanner without appearance enhancement. The clinical characteristics of the WORD dataset are listed in Table~\ref{tab:char_label}. Each CT volume consists of 159 to 330 slices of 512 $\times$ 512 pixels, with an in-plane resolution of 0.976 mm $\times$ 0.976 mm and slice spacing of 2.5 mm to 3.0 mm, indicating that the WORD dataset is a very high-resolution dataset. All scans of WORD dataset are exhaustively annotated with 16 anatomical organs, including the liver, spleen, kidney (L), kidney (R), stomach, gallbladder, esophagus, duodenum, colon, intestine, adrenal, rectum, bladder, head of the femur (L) and head of the femur (R). An example of image and annotation from the WORD dataset is shown in Figure~\ref{fig:show_data}. All images were anonymized and approved by the ethics committee to protect privacy where all clinical treatment details have been deleted. We randomly split WORD dataset into three parts: 100 scans (20115 slices) for training, 20 scans (4103 slices) for validation, and 30 scans (6277 slices) for testing. Figure~\ref{fig:data_dist} shows the volume distributions of all annotated organs. It shows that the extremely unbalanced distribution among large and small organs may bring some challenges to the segmentation task. At the same time, we further selected and annotated 20 CT scans from  LiTS~\citep{bilic2019liver} as an external evaluation set. These scans cover the whole abdominal region, each with 16 organ annotations.

\subsection{Professional data annotation}
Recently, the AbdomenCT-1K dataset~\citep{ma2021abdomenct} established an abdominal organ dataset using the pre-trained model for predictions and then refining by radiologists. At the same time, CT-ORG~\citep{rister2020ct} annotated the abdominal organ by using a semi-automatic tool firstly (ITK-SNAP~\citep{yushkevich2006user}) and then refining manually. However, these initial segmentation results could affect the annotator's decision, especial regarding low-contrast boundary regions. Differently from AbdomenCT-1K~\citep{ma2021abdomenct} and CT-ORG~\citep{rister2020ct}, all scans in the WORD dataset are annotated from scratch manually. A senior oncologist (with 7 years of experience) uses ITK-SNAP~\citep{yushkevich2006user} to delineate all organs slice-by-slice in axial view. {After that, an expert in oncology (more than 20 years of experience) checks and revises these annotations carefully and discusses them in cases of disagreement to produce consensus annotations and further ensure the annotation quality. Finally, these consensus labels are released and used for methods or clinical application development and evaluation. Note that all annotations and consensus discussions obey the radiation therapy delineation guideline published by Radiation Therapy Oncology Group (RTOG)\footnote{https://www.rtog.org/}. Here, we analyse the inconsistent ratio between the annotation of senior oncologist, the expert and their consensus annotations in Table~\ref{tab:revision_ratio}, suggesting that annotators' discrepancy is minor and their consensus labels are reliable.} In the annotation stage, each volume roughly takes 1.2-2.6 hours to annotate all 16 organs and further requires 0.4-1.0 hours to check, discuss and refine the annotation. The WORD dataset takes us around 15 months to collect, annotate and review, so we think it's precious and desirable to share with the medical image analysis community.

\begin{table*}[thp]
\centering
\normalsize
\scalebox{0.66}{\begin{tabular}{l|c|c|c|c|c|c|c|c|c|r}
\hline
\textbf{Method}                   & \textbf{nnUNet(2D)}       & \textbf{nnUNetV2(2D)}     & {ResUNet(2D)}      & \textbf{DeepLabV3+(2D)}   & \textbf{UNet++(2D)}       & \textbf{AttUNet(3D)}      & \textbf{nnUNet(3D)}       & \textbf{nnUNetV2(3D)}     & \textbf{UNETR(3D)}        & \textbf{CoTr(3D)} \\ 
\hline
\textbf{Liver}                  & {95.38$\pm$4.45}  & {96.19$\pm$2.16}  & {96.55$\pm$0.89}  & {96.21$\pm$1.34}  & {96.33$\pm$1.40}  & {96.00$\pm$1.01}  & {96.45$\pm$0.85}  & \textbf{96.59$\pm$6.10}  & {94.67$\pm$1.92}  & 95.58$\pm$1.59               \\  
\textbf{Spleen }          & {93.33$\pm$11.85} & {94.33$\pm$7.72}  & {95.26$\pm$2.84}  & {94.68$\pm$5.64}  & {94.64$\pm$4.22}  & {94.90$\pm$1.63}  & {95.98$\pm$0.89}  & \textbf{96.09$\pm$8.10}  & {92.85$\pm$3.03}  & 94.9$\pm$1.37                \\  
\textbf{Kidney (L) } & {90.05$\pm$19.35} & {91.29$\pm$18.15} & {95.63$\pm$1.20}  & {92.01$\pm$13.00} & {93.36$\pm$5.06}  & {94.65$\pm$1.38}  & {95.40$\pm$0.95}  & \textbf{95.63$\pm$9.20}  & {91.49$\pm$5.81}  & 93.26$\pm$3.07               \\  
\textbf{Kidney (R)}        & {89.86$\pm$19.56} & {91.20$\pm$17.22} & {95.84$\pm$1.16}  & {91.84$\pm$14.41} & {93.34$\pm$7.38}  & {94.7$\pm$2.78}   & {95.68$\pm$1.07}  & \textbf{95.83$\pm$9.00}  & {91.72$\pm$7.06}  & 93.63$\pm$3.01               \\  
\textbf{Stomach}          & {89.97$\pm$4.96}  & {91.12$\pm$3.60}  & {91.58$\pm$2.86}  & {91.16$\pm$3.07}  & {91.33$\pm$3.13}  & {91.15$\pm$2.74}  & \textbf{91.69$\pm$2.5}   & {91.57$\pm$3.05}  & {85.56$\pm$6.12}  & 89.99$\pm$4.49               \\  
\textbf{Gallbladder}      & {78.43$\pm$16.48} & {83.19$\pm$12.22} & {82.83$\pm$11.8}  & {80.05$\pm$17.92} & {81.21$\pm$12.24} & {81.38$\pm$10.95} & {83.19$\pm$8.81}  & \textbf{83.72$\pm$8.19}  & {65.08$\pm$19.63} & 76.4$\pm$16.48               \\  
\textbf{Esophagus}        & {78.08$\pm$13.99} & {77.79$\pm$13.51} & {77.17$\pm$14.68} & {74.88$\pm$14.69} & {78.36$\pm$12.84} & {76.87$\pm$15.12} & \textbf{78.51$\pm$12.22} & {77.36$\pm$13.66} & {67.71$\pm$13.46} & 74.37$\pm$14.92              \\  
\textbf{Pancreas}         & {82.33$\pm$6.5}   & {83.55$\pm$5.87}  & {83.56$\pm$5.60}  & {82.39$\pm$6.68}  & {84.43$\pm$6.77}  & {83.55$\pm$6.2}   & \textbf{85.04$\pm$5.78}  & {85.00$\pm$5.95}  & {74.79$\pm$9.31}  & 81.02$\pm$7.23               \\  
\textbf{Duodenum}         & {63.47$\pm$15.81} & {64.47$\pm$15.87} & {66.67$\pm$15.36} & {62.81$\pm$15.21} & {65.99$\pm$15.79} & {67.68$\pm$14.01} & \textbf{68.31$\pm$16.29} & {67.73$\pm$16.75} & {57.56$\pm$11.23} & 63.58$\pm$14.88              \\  
\textbf{Colon}                  & {83.06$\pm$8.32}  & {83.92$\pm$8.45}  & {83.57$\pm$8.69}  & {82.72$\pm$8.79}  & {83.22$\pm$8.98}  & {85.72$\pm$8.50}  & \textbf{87.41$\pm$7.38}  & {87.26$\pm$8.25}  & {74.62$\pm$11.5}  & 84.14$\pm$7.82               \\  
\textbf{Intestine}              & {85.6$\pm$4.08}   & {86.83$\pm$4.02}  & {86.76$\pm$3.56}  & {85.96$\pm$4.02}  & {86.37$\pm$4.01}  & {88.19$\pm$3.34}  & {89.3$\pm$2.75}   & \textbf{89.37$\pm$3.11}  & {80.4$\pm$4.59}   & 86.39$\pm$3.51               \\  
\textbf{Adrenal}                & {69.9$\pm$11.07}  & {70.0$\pm$11.86}  & {70.9$\pm$10.12}  & {66.82$\pm$10.81} & {71.04$\pm$10.65} & {70.23$\pm$9.31}  & {72.38$\pm$8.98}  & \textbf{72.98$\pm$8.09}  & {60.76$\pm$8.32}  & 69.06$\pm$9.26               \\  
\textbf{Rectum}                 & {81.66$\pm$6.64}  & {81.49$\pm$7.37}  & {82.16$\pm$6.73}  & {81.85$\pm$6.67}  & {81.44$\pm$6.7}   & {80.47$\pm$5.44}  & \textbf{82.41$\pm$4.9}   & {82.32$\pm$5.26}  & {74.06$\pm$8.03}  & 80.0$\pm$5.4                 \\  
\textbf{Bladder}                & {90.49$\pm$14.73} & {90.15$\pm$16.85} & {91.0$\pm$13.5}   & {90.86$\pm$14.07} & {92.09$\pm$11.53} & {89.71$\pm$15.00} & \textbf{92.59$\pm$8.27}  & {92.11$\pm$9.75}  & {85.42$\pm$18.17} & 89.27$\pm$18.28              \\  
\textbf{Head of Femur (L)}       & {93.28$\pm$5.31}  & {93.28$\pm$5.12}  & \textbf{93.39$\pm$5.11}  & {92.01$\pm$4.76}  & {93.38$\pm$5.12}  & {91.90$\pm$4.39}  & {91.99$\pm$4.72}  & {92.56$\pm$4.19}  & {89.47$\pm$6.4}   & 91.03$\pm$4.81               \\  
\textbf{Head of Femur (R)}       & {93.78$\pm$4.38}  & \textbf{93.93$\pm$4.29}  & {93.88$\pm$4.30}  & {92.29$\pm$4.01}  & {93.88$\pm$4.21}  & {92.43$\pm$3.68}  & {92.74$\pm$3.63}  & {92.49$\pm$4.03}  & {90.17$\pm$4.0}   & 91.87$\pm$3.32               \\  
\hline
\textbf{Mean}                   & {84.92$\pm$5.39}  & {85.80$\pm$5.27}  & {86.67$\pm$4.81}  & {84.91$\pm$5.05}  & {86.28$\pm$3.96}  & {86.21$\pm$4.78}  & \textbf{87.44 $\pm$4.33} & {87.41$\pm$4.57}  & {79.77$\pm$4.92}  & 84.66$\pm$5.45               \\ 
\hline
\end{tabular}}
\caption{{Performance comparison ($DSC~(\%)$) of 16 abdominal organs segmentation using ten recent segmentation methods.}}
\label{tab:full_dsc}
\end{table*}

\begin{table*}[thp]
\centering
\normalsize
\scalebox{0.66}{\begin{tabular}{l|c|c|c|c|c|c|c|c|c|r}
\hline
\textbf{Method}              & 
\textbf{nnUNet(2D)}       &
\textbf{nnUNetV2(2D)}     &
\textbf{ResUNet(2D)}      & 
\textbf{DeepLabV3+(2D)}   & 
\textbf{UNet++(2D)}       & 
\textbf{AttUNet(3D)}      &
\textbf{nnUNet(3D)}       & 
\textbf{nnUNetV2(3D)}     &
\textbf{UNETR(3D)}        &\textbf{CoTr(3D)} \\ 
\hline
\textbf{Liver}             & {7.94$\pm$17.23}  & {7.34$\pm$16.48}  & {4.64$\pm$7.37}   & {6.81$\pm$18.3}   & {11.77$\pm$22.17} & {3.61$\pm$1.75}   & {3.31$\pm$1.38}   & \textbf{3.17$\pm$0.51}   & {8.36$\pm$14.13}  & 7.47$\pm$12.18               \\ 
\textbf{Spleen}            & {14.46$\pm$41.27} & {9.53$\pm$33.84}  & {8.7$\pm$30.11}   & {8.93$\pm$33.61}  & {9.39$\pm$32.14}  & {2.74$\pm$1.61}   & {2.15$\pm$0.5}    & \textbf{2.12$\pm$0.47}   & {14.84$\pm$34.62} & 8.14$\pm$24.43               \\ 
\textbf{Kidney (L)}        & {10.53$\pm$29.43} & {10.33$\pm$29.52} & {5.4$\pm$15.85}   & {10.4$\pm$29.39}  & {13.09$\pm$29.75} & {6.28$\pm$19.19}  & {6.07$\pm$19.38}  & \textbf{2.46$\pm$0.7}    & {23.37$\pm$39.28} & 16.42$\pm$27.79              \\ 
\textbf{Kidney (R)}   & {10.73$\pm$28.49} & {10.85$\pm$28.41} & {2.47$\pm$0.97}   & {10.02$\pm$28.0}  & {21.84$\pm$42.61} & {2.86$\pm$1.46}   & {2.35$\pm$0.81}   & {2.24$\pm$0.47}   & {7.9$\pm$19.08}   & 12.79$\pm$29.76              \\ 
\textbf{Stomach}     & {19.04$\pm$20.82} & {13.97$\pm$12.08} & {9.98$\pm$6.62}   & {11.01$\pm$8.45}  & {15.4$\pm$21.44}  & \textbf{8.23$\pm$6.07}   & {8.47$\pm$5.96}   & {9.47$\pm$7.61}   & {19.25$\pm$23.19} & 10.26$\pm$9.49               \\ 
\textbf{Gallbladder} & {8.9$\pm$10.33}   & {7.91$\pm$8.67}   & {9.48$\pm$12.97}  & {7.36$\pm$9.43}   & {14.68$\pm$28.48} & \textbf{5.11$\pm$3.41}   & {5.24$\pm$5.3}    & {6.04$\pm$5.63}   & {12.72$\pm$15.39} & 11.32$\pm$15.57              \\ 
\textbf{Esophagus}   & {6.9$\pm$9.72}    & {6.7$\pm$7.8}     & {6.7$\pm$7.6}     & {7.6$\pm$8.45}    & {5.85$\pm$3.93}   & \textbf{5.35$\pm$3.79}   & {5.49$\pm$4.34}   & {5.83$\pm$4.64}   & {9.31$\pm$8.41}   & 6.29$\pm$4.53                \\ 
\textbf{Pancreas}    & {7.92$\pm$7.34}   & {7.82$\pm$6.76}   & {7.82$\pm$7.15}   & {7.67$\pm$7.1}    & {7.5$\pm$8.45}    & {6.96$\pm$7.39}   & \textbf{6.84$\pm$7.9}    & {6.87$\pm$7.86}   & {10.66$\pm$8.56}  & 8.88$\pm$10.61               \\ 
\textbf{Duodenum}    & {25.18$\pm$18.39} & {23.29$\pm$14.39} & {21.79$\pm$12.83} & {21.61$\pm$13.88} & {23.67$\pm$13.8}  & {21.61$\pm$12.86} & {21.3$\pm$14.22}  & \textbf{21.15$\pm$14.26} & {25.15$\pm$21.96} & 24.83$\pm$15.47              \\ 
\textbf{Colon}             & {15.56$\pm$12.97} & {15.68$\pm$14.0}  & {17.41$\pm$15.22} & {15.95$\pm$14.07} & {16.97$\pm$13.92} & {10.21$\pm$12.87} & \textbf{9.99$\pm$13.17}  & {10.42$\pm$14.27} & {20.32$\pm$14.37} & 12.41$\pm$12.75              \\ 
\textbf{Intestine}         & {10.46$\pm$6.24}  & {8.96$\pm$4.83}   & {9.54$\pm$7.2}    & {9.57$\pm$5.21}   & {10.06$\pm$6.01}  & {5.68$\pm$3.93}   & \textbf{5.14$\pm$3.68}   & {5.27$\pm$4.29}   & {12.62$\pm$7.63}  & 7.96$\pm$5.58                \\ 
\textbf{Adrenal}           & {6.06$\pm$3.99}   & {6.42$\pm$4.3}    & {6.67$\pm$4.59}   & {7.14$\pm$4.8}    & {7.14$\pm$4.97}   & {5.98$\pm$4.01}   & {5.46$\pm$4.04}   & \textbf{5.43$\pm$3.82}   & {8.73$\pm$5.3}    & 6.76$\pm$6.99                \\ 
\textbf{Rectum}            & \textbf{10.62$\pm$5.5}   & {11.15$\pm$7.33}  & {10.62$\pm$6.52}  & {10.96$\pm$6.94}  & {11.54$\pm$8.13}  & {11.67$\pm$6.37}  & {11.57$\pm$6.95}  & {12.39$\pm$8.12}  & {12.79$\pm$6.38}  & 11.26$\pm$6.06               \\ 
\textbf{Bladder}           & {5.88$\pm$7.21}   & {4.97$\pm$5.26}   & {5.02$\pm$6.17}   & {5.14$\pm$6.22}   & {5.06$\pm$6.56}   & {4.83$\pm$4.66}   & \textbf{3.68$\pm$2.23}   & {4.17$\pm$3.6}    & {14.71$\pm$40.82} & 14.34$\pm$43.85              \\ 
\textbf{Head of Femur (L)}  & {6.56$\pm$8.09}   & \textbf{6.54$\pm$8.13}   & {6.56$\pm$8.3}    & {7.62$\pm$7.93}   & {6.66$\pm$8.22}   & {6.93$\pm$6.27}   & {35.18$\pm$88.78} & {17.05$\pm$62.15} & {38.11$\pm$98.44} & 19.42$\pm$70.83              \\ 
\textbf{Head of Femur (R)}  & {5.89$\pm$7.55}   & \textbf{5.74$\pm$6.76}   & {5.98$\pm$7.2}    & {7.02$\pm$6.76}   & {16.92$\pm$63.02} & {6.06$\pm$4.78}   & {33.03$\pm$82.19} & {27.29$\pm$81.62} & {38.62$\pm$99.75} & 26.78$\pm$78.4               \\ 
\hline
\textbf{Mean}              & {10.79$\pm$10.29} & {9.88$\pm$9.16}   & {8.6$\pm$6.47}    & {9.67$\pm$9.06}   & {12.35$\pm$15.87} & \textbf{7.13$\pm$4.68}   & {10.33$\pm$26.65} & {8.84$\pm$ 22.63} & {17.34$\pm$28.8}  & 12.83$\pm$21.96              \\ 
\hline
\end{tabular}}
\caption{{Performance comparison ($HD_{95}~(mm)$) of 16 abdominal organs segmentation using ten recent segmentation methods.}}
\label{tab:full_hd}
\end{table*}

\subsection{Potential research topics}
{We can conduct many essential research topics on medical image segmentation/detection methods and clinical application with the large and carefully annotated abdominal multi-organ dataset. Besides, there are some challenges in the WORD, including the imbalanced sample among large and small size organs, the high resolution, and the complex anatomical structure. It can be used to develop or evaluate clinical application, as it is very desirable to develop a tool or software to assist oncologists in delineating organs quickly and accurately. The WORD dataset also can be employed in general algorithm research, such as fully-/semi-/weakly-supervised learning, domain adaption/generalization, partially label and lifelong learning, etc. Here, we roughly summarized the potential research topics as follows.}

\subsubsection{Fully-supervised abdominal organ segmentation and generalization}
Fully supervised learning~\citep{isensee2021nnu} aims to efficiently utilize the labelled data to achieve good results and solve the challenges of imbalanced distributions and complex structures. It is a fundamental topic and has been studied for many years. Here, we presented a new abdominal organ segmentation dataset to boost abdominal organ segmentation algorithm research, evaluation, and comparison. Afterwards, we further build a publicly external evaluation dataset from LiTS~\citep{bilic2019liver} for segmentation models' generalization evaluation. In addition, the WORD dataset and the external dataset can be used to develop clinical application or clinically applicable evaluations.
\subsubsection{Abdominal organ segmentation with low computational cost and high speed}For 3D abdominal CT scans, the inference stage always takes much time and GPU memory due to the high dimension and resolution. To deal with this issue, inference-efficient model~\citep{feng2021resolution} is proposed to achieve the trade-off between high-performance and low inference cost. However, very few works have been studied to accelerate the inference of 3D medical image segmentation tasks~\citep{tang2021high}. Recently, knowledge distillation has achieved success in several 2D natural image recognition tasks, which may have the potential to handle the 3D medical image segmentation tasks~\citep{mishra2017apprentice}.
\subsubsection{Abdominal organ segmentation with low annotation cost}Pixel-wise abdominal organ annotation is very expensive, requiring clinical experience and much time. Recently, annotation-efficient learning~\citep{luo2020semi,zhou2019semi,luo2021ctbct,media2022urpc} has been introduced to reduce the labeling cost and improve the network generalization ability by semi-/weakly-supervised learning, domain adaptation strategies, etc. These strategies have been scorching topics and show the potential to reduce annotation cost by utilizing unlabeled data or sparse annotations. Reducing the annotation cost for accurate abdominal organ segmentation is desirable, as it can accelerate the model development and reduce cost. In this work, we pay more attention to evaluating weakly supervised methods to reduce labeling cost.

\section{Experiments and Analyses}
\subsection{Implementations and metrics}
In this work, all methods are implemented, trained, and tested by PyTorch 1.8~\citep{paszke2019pytorch} on a cluster with eight NVIDIA GTX1080TI GPUs. We choose the powerful nnUNet\footnote{https://github.com/MIC-DKFZ/nnUNet}~\citep{isensee2021nnu} as our baseline for fair comparisons. nnUNet is a self-configuration segmentation framework without needing any manual effort for data processing, training planning (network architectures and parameters setting, etc.), and post-processing, and has won more than 19 medical segmentation challenges~\citep{isensee2021nnu}. Due to that the nnUNet just provides implementation of the vanilla UNet network, we further adapt it to support more network architectures. Note that we use the public implementations of the compared methods\footnote{https://github.com/qubvel/segmentation\_models. pytorch}. We employ the default settings of nnUNet as our experimental settings, where the batch size is 2 for 3D methods and 12 for 2D methods, the total epoch is 1000, and the loss function is a combination of cross-entropy and dice loss. All the models are trained and tested based on the default settings, except that we don't use the test time augmentation, as each model needs more than six GPU days to train, and each volume takes more than five minutes to infer. We use two widely-used metrics to measure the segmentation quality in this work: 1) Dice similarity coefficient ($DSC$) is used to evaluate the pixel-wise overlap between the ground truth and prediction; 2) 95\% Hausdorff Distance ($HD_{95}$) that measures distance difference between the boundaries of the ground truth and prediction. The implementations of $DSC$ and $HD_{95}$ are available\footnote{https://github.com/loli/medpy}.

\begin{figure*}[ht]
    \centering
    \includegraphics[width=0.99\textwidth]{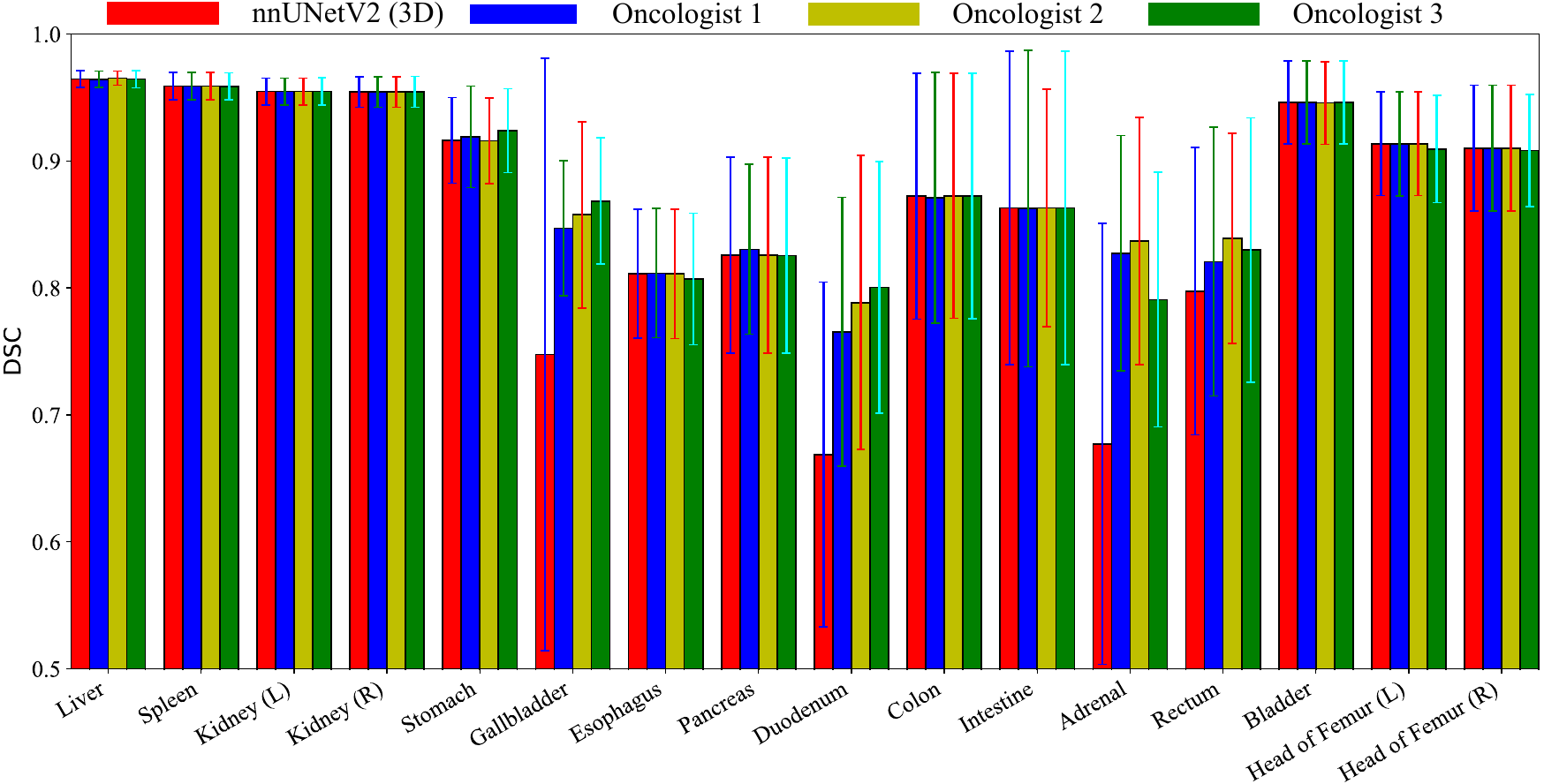}
    \caption{User study based on three junior oncologists independently, each of them comes from a different hospital.}
    \label{fig:revision_degree}
\end{figure*}

\subsection{Fully-supervised abdominal organ segmentation}
We first evaluate some existing state-of-the-art (SOTA) methods on the WORD dataset. Then, we further evaluate the gap between the deep network and three oncologists. Finally, we investigate the domain shift between the WORD dataset and three public datasets (BTCV~\citep{landman2017multi}, TCIA~\citep{roth2015deeporgan}, and LiTS~\citep{bilic2019liver}).
\subsubsection{Evaluations of SOTA methods on the WORD}
{For deep learning-based clinical application, fully supervised learning is one of the most basic and popular solutions, especially in automatic multi-organ delineation systems. In this work, we investigate several existing SOTA methods on the WORD dataset, including convolutional neural networks-based networks,  nnUNet~\citep{isensee2021nnu} and its variations (both 2D and 3D), ResUNet~\citep{diakogiannis2020resunet}, DeepLabV3+~\citep{chen2018encoder}, UNet++~\citep{zhou2019unet++} and Attention-UNet (AttUNet)~\citep{oktay2018attention}, and transformer-based architectures, CoTr~\citep{xie2021cotr} and UNETR~\citep{hatamizadeh2022unetr}. The quantitative segmentation results in term of $DSC$ and $HD_{95}$ are presented in Table~\ref{tab:full_dsc} and Table~\ref{tab:full_hd}, respectively. It can be observed that all CNN-based methods outperform transformer-based CoTr~\citep{xie2021cotr} and UNETR~\citep{hatamizadeh2022unetr}. Moreover, the results further show that all SOTA methods can achieve very promising results ($DSC$ $\textgreater$ 85\%) on large organs, such as the liver, spleen, kidney, stomach, bladder, and head of the femur. It has also proven that the large organ segmentation task is a well-solved problem if there are enough high-quality annotated samples. But for the gallbladder, pancreas, and rectum segmentation, almost all methods get poor results, where $DSC$ $\textless$ 85\% and $HD_{95}$ $\textgreater$ 10$mm$. In addition, the the segmentation results of esophagus, duodenum and adrenal are extremely bad, where almost all methods achieve $DSC$ $\textless$ 70\%. All the above results show that the segmentation of small organs remains challenging and needs more attention to be paid. However, few works and researchers have focused on solving these challenging tasks. One of the critical reasons is lacking large-scale and publicly available datasets and benchmarks. To alleviate these challenges, we build the WORD dataset and corresponding benchmarks to boost research in the medical image computing community.}

\subsubsection{User study by three oncologists}
Then, we employ a comprehensive user study to measure the gap between the network and three oncologists. Following the general workflow of deep learning-assisted organs delineation systems~\citep{chen2021deep}, we invite three junior oncologists (with 3 years of experience) from three different hospitals to revise model-generated predictions independently until the results are clinically acceptable. We randomly selected 20 predictions produced by nnUNetV2 (3D) for the user study and calculated the revised results. The quantitative comparison in terms of $DSC$ between the nnUNet predictions and three oncologists' revised results are presented in Figure~\ref{fig:revision_degree}. For organs with large size and clear boundary, the deep network can produce promising results that are very close to clinically applicable with just a few revisions. However, there is a massive gap between the deep network and junior oncologists in small organ segmentation. It indicates that the deep network has the potential to reduce the burden of oncologists in annotating large organs. In the future, combining the user interaction with the deep network may help further to reduce the burden of delineating small organs and accelerate the clinical workflow~\citep{luo2021mideepseg,wang2018deepigeos}.

\begin{table}[ht]
\centering
\normalsize
\scalebox{0.46}{
\begin{tabular}{l|c|c|c|c|c|r}
\hline
\textbf{DataSet}                                   & \multicolumn{2}{c|}{\textbf{TCIA}}                                                    & \multicolumn{2}{c|}{\textbf{BTCV}}                                                    & \multicolumn{2}{c}{\textbf{LiTS}}                               \\   
\hline
\textbf{Method}                 & \textbf{nnUNetV2 (2D)}   & \textbf{nnUNetV2 (3D)}   & \textbf{nnUNetV2 (2D)}   & \textbf{nnUNetV2 (3D)}   & \textbf{nnUNetV2 (2D)}   & \textbf{nnUNetV2 (3D)}   \\ 
\hline
\textbf{Liver}                  & {91.27$\pm$3.71}  & \textbf{92.59$\pm$3.72}  & {86.63$\pm$7.79}  & \textbf{93.36$\pm$5.75}  & {92.37$\pm$3.48}  & \textbf{94.2$\pm$1.41}   \\    
\textbf{Spleen }          & {85.47$\pm$13.97} & \textbf{86.31$\pm$12.57} & {72.43$\pm$19.36} & \textbf{88.89$\pm$11.78} & {84.73$\pm$14.43} & \textbf{95.28$\pm$5.44}  \\    
\textbf{Kidney (L) } & {80.24$\pm$13.4}  & \textbf{91.44$\pm$5.13}  & {64.57$\pm$23.94} & \textbf{87.36$\pm$16.25} & {79.23$\pm$19.24} & \textbf{96.33$\pm$3.83}  \\    
\textbf{Kidney (R)}        & {-}               & {-}               & {37.32$\pm$33.71} & \textbf{56.22$\pm$42.99} & {78.51$\pm$22.39} & \textbf{96.07$\pm$4.84}  \\    
\textbf{Stomach}          & {54.32$\pm$21.73} & \textbf{73.37$\pm$17.35} & {51.73$\pm$22.74} & \textbf{78.52$\pm$18.54} & {68.46$\pm$15.88} & 86.43$\pm$13.68 \\    
\textbf{Gallbladder}      & {54.0$\pm$32.36}  & \textbf{78.49$\pm$18.33} & {40.38$\pm$32.44} & \textbf{63.82$\pm$32.05} & {49.05$\pm$32.97} & \textbf{60.15$\pm$34.37} \\    
\textbf{Esophagus}        & {54.62$\pm$21.38} & \textbf{61.22$\pm$18.85} & {47.1$\pm$18.37}  & \textbf{62.53$\pm$15.28} & {84.77$\pm$5.04}  & \textbf{87.13$\pm$6.8}   \\    
\textbf{Pancreas}         & {51.38$\pm$21.59} & \textbf{68.53$\pm$15.25} & {49.64$\pm$17.59} & \textbf{73.64$\pm$13.15} & {59.59$\pm$12.78} & \textbf{89.43$\pm$4.16}  \\    
\textbf{Duodenum}         & {33.15$\pm$18.92} & \textbf{51.14$\pm$15.78} & {24.19$\pm$14.78} & \textbf{56.19$\pm$15.62} & {45.01$\pm$15.52} & \textbf{76.45$\pm$6.4}   \\    
\textbf{Colon}                  & {-}               & {-}               & {-}               & {-}               & {75.42$\pm$14.65} & \textbf{87.54$\pm$8.32}  \\    
\textbf{Intestine}              & {-}               & {-}               & {-}               & {-}               & {64.35$\pm$10.77} & 83.6$\pm$6.87   \\    
\textbf{Adrenal}                & {-}               & {-}               & {17.01$\pm$20.72} &\textbf{41.72$\pm$32.5}  & {62.86$\pm$14.16} & \textbf{85.89$\pm$4.49}  \\    
\textbf{Rectum}                 & {-}               & {-}               & {-}               & {-}               & {68.93$\pm$24.83} & \textbf{80.61$\pm$19.57} \\    
\textbf{Bladder}                & {-}               & {-}               & {-}               & {-}               & {91.63$\pm$5.94}  & \textbf{92.88$\pm$7.43}  \\    
\textbf{Head of Femur (L)}       & {-}               & {-}               & {-}               & {-}               & {93.26$\pm$2.37}  & \textbf95.55$\pm$2.63  \\    
\textbf{Head of Femur (R)}       & {-}               & {-}               & {-}               & {-}               & {92.73$\pm$3.12}  & \textbf{94.98$\pm$3.51}  \\    
\hline
\textbf{Mean}                   & {63.06$\pm$7.79}  & \textbf{75.39$\pm$5.50}  & {49.10$\pm$7.34}  & \textbf{70.23$\pm$10.97} & {74.43$\pm$8.29}  & \textbf{87.66$\pm$7.98}  \\
\hline
\end{tabular}

}
\caption{{The segmentation result ($DSC~(\%)$) of the BTCV~\citep{landman2017multi}, TCIA~\citep{roth2015deeporgan} and LiTS~\citep{bilic2019liver} using the pre-trained (on the WORD) nnUNetV2 (2D/3D).}}
\label{tab:gene_tcia_dice}
\end{table} 

\begin{table}[ht]
\centering
\normalsize
\scalebox{0.46}{
\begin{tabular}{l|c|c|c|c|c|r}
\hline
\textbf{DataSet}                                   & \multicolumn{2}{c|}{\textbf{TCIA}}                                                    & \multicolumn{2}{c|}{\textbf{BTCV}}                                                    & \multicolumn{2}{c}{\textbf{LiTS}}                               \\   
\hline
\textbf{Method}                 & \textbf{nnUNetV2 (2D)}   & \textbf{nnUNetV2 (3D)}   & \textbf{nnUNetV2 (2D)}   & \textbf{nnUNetV2 (3D)}   & \textbf{nnUNetV2 (2D)}   & \textbf{nnUNetV2 (3D)}   \\ 
\hline
\textbf{Liver} & {29.39$\pm$20.67} & \textbf{17.11$\pm$24.27} & {44.73$\pm$36.15}  & \textbf{13.96$\pm$23.73} & {32.28$\pm$27.58}  & \textbf{11.57$\pm$9.76}   \\    
\textbf{Spleen }          & {87.32$\pm$69.61} & \textbf{31.11$\pm$45.85} & {128.28$\pm$59.62} & \textbf{37.46$\pm$67.09} & {90.59$\pm$63.72}  & \textbf{12.13$\pm$21.25} \\    
\textbf{Kidney (L) } & {92.93$\pm$49.56} & \textbf{15.44$\pm$39.47} & {107.0$\pm$40.9}   & \textbf{20.34$\pm$38.08} & {96.44$\pm$51.64}  & \textbf{12.15$\pm$29.13}  \\    
\textbf{Kidney (R)}        & {-}               & {-}               & {71.58$\pm$43.5}   & \textbf{27.8$\pm$28.6}   & {69.78$\pm$64.44}  & \textbf{10.16$\pm$34.59}  \\    
\textbf{Stomach}          & {43.17$\pm$23.49} & \textbf{26.75$\pm$22.28} & {65.47$\pm$42.91}  & \textbf{31.42$\pm$40.64} & {47.33$\pm$31.66}  & \textbf{17.87$\pm$28.48} \\    
\textbf{Gallbladder}      & {34.94$\pm$40.91} & \textbf{7.65$\pm$11.62}  & {53.34$\pm$61.11}  & \textbf{16.69$\pm$21.8}  & {28.22$\pm$38.44}  & \textbf{18.63$\pm$22.57}  \\    
\textbf{Esophagus}        & {17.41$\pm$9.98}  & \textbf{15.85$\pm$9.66}  & {22.25$\pm$15.17}  & \textbf{18.81$\pm$18.02} & \textbf{4.44$\pm$2.17}    & 5.91$\pm$8.08    \\    
\textbf{Pancreas}         & {31.37$\pm$8.59}  & \textbf{15.59$\pm$13.67} & {30.15$\pm$15.12}  & \textbf{13.12$\pm$18.01} & {27.21$\pm$11.79}  & \textbf{4.93$\pm$3.66}    \\    
\textbf{Duodenum}         & {34.09$\pm$14.13} & \textbf{35.07$\pm$18.54} & {50.97$\pm$26.19}  & \textbf{29.0$\pm$15.09}  & {25.97$\pm$9.99}   & \textbf{15.75$\pm$9.52}   \\    
\textbf{Colon}                  & {-}               & {-}               & {-}                & {-}               & {31.89$\pm$14.66}  & \textbf{24.82$\pm$45.11}  \\    
\textbf{Intestine}              & {-}               & {-}               & {-}                & {-}               & {31.15$\pm$10.63}  & \textbf{12.11$\pm$6.55}   \\    
\textbf{Adrenal}                & {-}               & {-}               & {26.28$\pm$37.47}  & \textbf{5.29$\pm$9.75}   & {6.11$\pm$3.27}    & \textbf{2.22$\pm$0.72}    \\    
\textbf{Rectum}                 & {-}               & {-}               & {-}                & {-}               & {44.38$\pm$111.99} & \textbf{10.27$\pm$9.8}    \\    
\textbf{Bladder}                & {-}               & {-}               & {-}                & {-}               & \textbf{33.46$\pm$117.57} & 54.35$\pm$149.16 \\    
\textbf{Head of Femur (L)}       & {-}               & {-}               & {-}                & {-}               & {32.35$\pm$115.82} & \textbf{30.88$\pm$116.82} \\    
\textbf{Head of Femur (R)}       & {-}               & {-}               & {-}                & {-}               & {59.42$\pm$161.8}  & \textbf{58.08$\pm$162.04} \\  
\hline
\textbf{Mean}                   & {46.33$\pm$20.34} & \textbf{20.57$\pm$12.30} & {60.00$\pm$15.01}  & \textbf{21.39$\pm$15.96} & {41.31$\pm$47.99}  & \textbf{18.86$\pm$50.87}  \\
\hline
\end{tabular}

}
\caption{{The segmentation result ($HD_{95}~(mm)$) of the BTCV~\citep{landman2017multi}, TCIA~\citep{roth2015deeporgan} and LiTS~\citep{bilic2019liver} using the pre-trained (on the WORD) nnUNetV2 (2D/3D).}}
\label{tab:gene_tcia_hd95}
\end{table} 

\begin{figure}[t]
    \centering
    \includegraphics[width=0.49\textwidth]{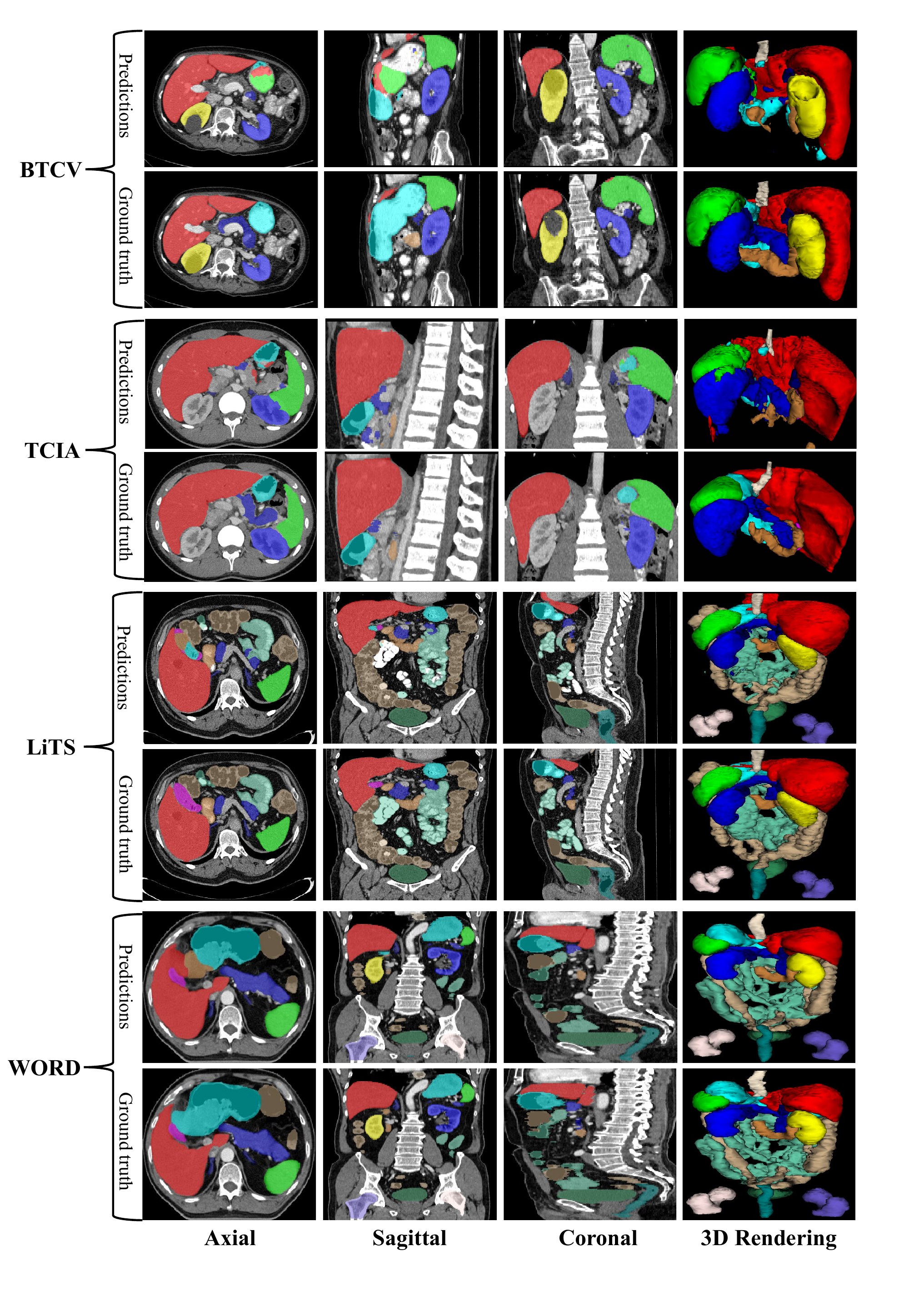}
    \caption{{Visual comparison of segmentation performance on four different datasets. All predictions were produced by the nnUNetV2 (3D) pre-trained on the WORD.
    }}
    \label{fig:generailization_vis}
\end{figure}

\begin{figure}[t]
    \centering
    \includegraphics[width=0.495\textwidth]{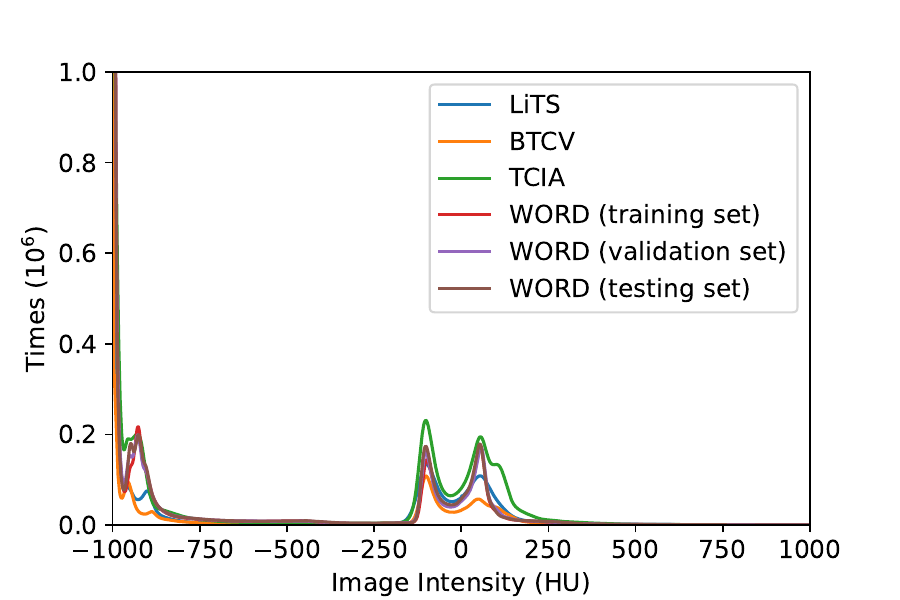}
    \caption{Intensity distributions comparison of LiTS, BTCV, TCIA and WORDs. HU means Hounsfield Unit.}
    \label{fig:intensity_histogram}
\end{figure}

\subsubsection{Generalization on BTCV, TCIA and LiTS}
We further investigate the domain shift between the WORD dataset and three widely-used public datasets BTCV~\citep{landman2017multi}, TCIA~\citep{roth2015deeporgan} and LiTS~\citep{bilic2019liver}. The differences between the WORD dataset and BTCV~\citep{landman2017multi}, TCIA~\citep{roth2015deeporgan} and LiTS~\citep{bilic2019liver} lie in 1) coming from different centres/scanners/countries;  2) suffering from different diseases; 3) with different phase/contrast enhancement; 4) with different voxel spacing; 5) annotating by different oncologists/radiologists. All of them could affect the generalizability of the deep network and further limit clinical practice. In this work, we use the pre-trained model on the WORD dataset to infer the samples from BTCV~\citep{landman2017multi} (47 scans), TCIA~\citep{roth2015deeporgan} (43 scans) and LiTS~\citep{bilic2019liver} (20 scans) to estimate the domain gap. Table~\ref{tab:gene_tcia_dice} and Table~\ref{tab:gene_tcia_hd95} list the results of $DSC$ and $HD_{95}$, respectively. Here, we just consider the official annotated organs of the BTCV~\citep{landman2017multi} and TCIA~\citep{roth2015deeporgan} datasets. It can be found that there are very significant domain shifts between WORD dataset and BTCV~\citep{landman2017multi}, TCIA~\citep{roth2015deeporgan} datasets, as the pre-trained nnUNet on the WORD dataset performs very worse on the BTCV~\citep{landman2017multi} and TCIA~\citep{roth2015deeporgan}. For the LiTS~\citep{bilic2019liver} dataset, the nnUNetV2(3D) achieves very encouraging results, even better than the results of WORD. But the nnUNetV2(2D) still performs badly, which may be caused by that the nnUNetV2(2D) did not consider the relationship between neighbouring slices. It also indicates that the model generalization for the multi-site abdominal organ task is not a solved problem. Figure~\ref{fig:generailization_vis} shows some segmentation results of different datasets. These results are generated by a pre-trained nnUNetV2(3D) on the WORD. It can be observed that the results of TCIA and BTCV are inaccurate, which indicates that there is a significant domain gap between TCIA/BTCV and WORD. In contrast, the result of LiTS is better and more promising; the reason may be the domain gap between LiTS and WORD dataset is minor. {{In addition, we further analyse the intensity distributions between LiTS~\citep{bilic2019liver}, BTCV~\citep{landman2017multi}, TCIA~\citep{roth2015deeporgan}, and WORD in Figure~\ref{fig:intensity_histogram}. It shows there are bigger intensity distribution gaps between WORD dataset and BTCV/TCIA than LiTS, which conforms to segmentation results.}}

\begin{figure}[t]
    \centering
    \includegraphics[width=0.49\textwidth]{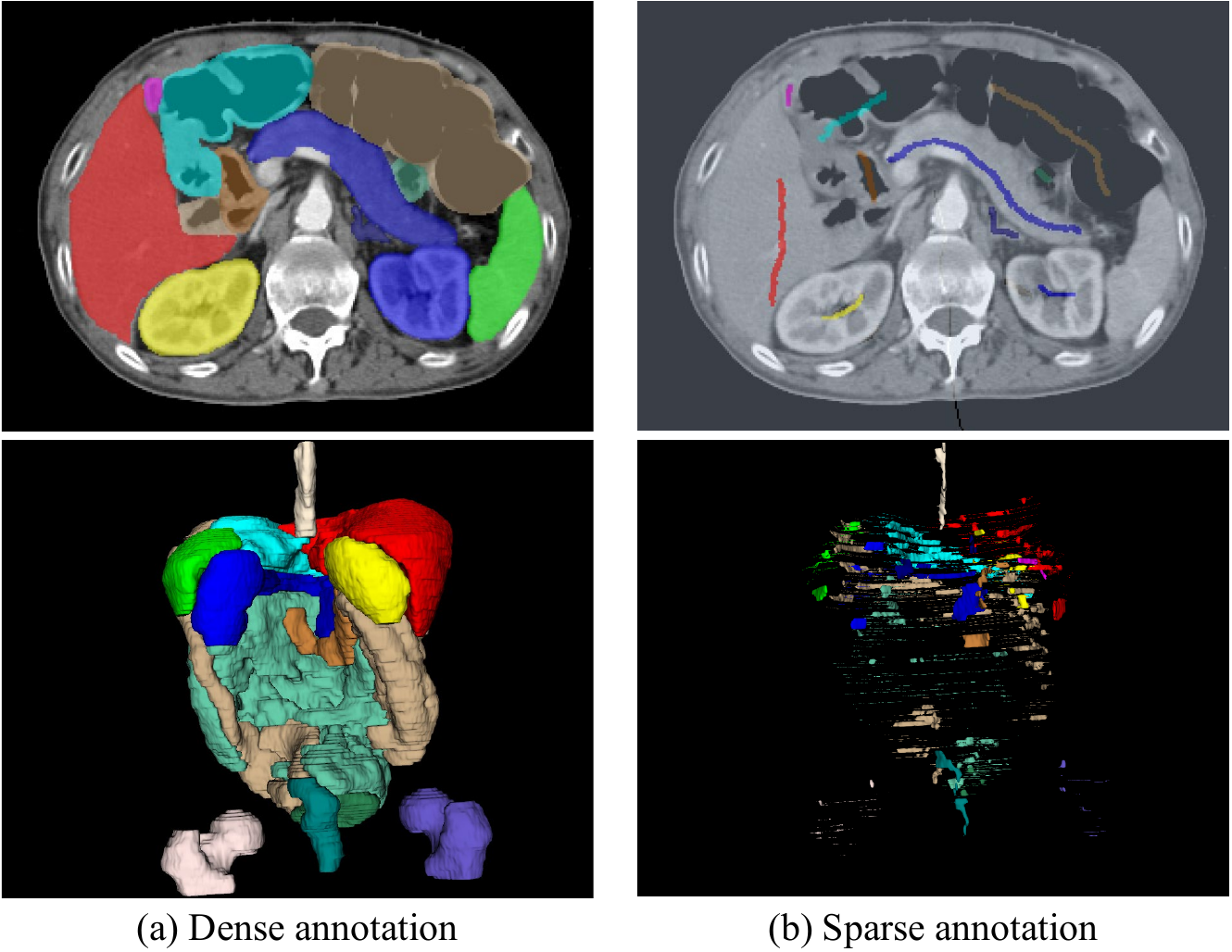}
    \caption{Different types of medical image annotation, the first and second rows show the visualization in 2D and 3D spaces, respectively.
    }
    \label{fig:annotation_diff}
\end{figure}

\begin{table}[thp]
\centering
\normalsize
\scalebox{0.44}{
\begin{tabular}{l|c|cc|cc|cr}
\hline
\textbf{Method} & \textbf{nnUNetV2(3D)}& \textbf{ESPNet}           & \textcolor{red}{\textbf{ESPNet+KD}}        &\textbf{DMFNet}           & \textbf{DMFNet+KD}        & \textbf{LCOVNet}          & \textbf{LCOVNet+KD} \\
\hline
\textbf{Liver}                  & {96.59$\pm$6.10}                                                                           & {76.47$\pm$19.12} & \textcolor{red}{{95.64$\pm$0.9}}  & {95.8$\pm$0.79}   & \textbf{95.96$\pm$0.76}  & {95.37$\pm$1.2}   & 95.89$\pm$0.58                 \\  
\textbf{Spleen }          & {96.09$\pm$8.10}                                                                           & {84.54$\pm$19.81} & \textcolor{red}{{93.9$\pm$3.22}}  & {94.25$\pm$2.15}  & {94.64$\pm$2.53}  & {94.6$\pm$2.08}   & \textbf{95.4$\pm$2.14}                  \\  
\textbf{Kidney (L) } & {95.63$\pm$9.20}                                                                           & {85.23$\pm$20.96} & \textcolor{red}{{92.24$\pm$6.04}}  & {94.72$\pm$0.97}  & {94.7$\pm$1.01}   & {94.78$\pm$1.18}  & \textbf{95.17$\pm$1.13}                 \\  
\textbf{Kidney (R)}        & {95.83$\pm$9.00}                                                                           & {89.65$\pm$15.74} & \textcolor{red}{{94.39$\pm$1.61}}  & {95.0$\pm$1.12}   & {94.96$\pm$1.19}  & {95.08$\pm$1.14}  & \textbf{95.78$\pm$0.84}                 \\  
\textbf{Stomach}          & {91.57$\pm$3.05}                                                                           & {82.87$\pm$12.06} & \textcolor{red}{{87.37$\pm$7.26}}  & {89.69$\pm$3.43}  & {89.88$\pm$5.24}  & {90.06$\pm$3.32}  & \textbf{90.86$\pm$3.82}                 \\  
\textbf{Gallbladder}      & {83.72$\pm$8.19}                                                                           & {49.02$\pm$29.45} & \textcolor{red}{{67.19$\pm$24.84}} & {77.12$\pm$17.7}  & \textbf{79.84$\pm$11.81} & {75.48$\pm$13.46} & 78.87$\pm$11.8                 \\  
\textbf{Esophagus}        & {77.36$\pm$13.66}                                                                          & {59.46$\pm$23.21} & \textcolor{red}{{67.91$\pm$14.27}} & {74.41$\pm$12.08} & {74.1$\pm$14.81}  & \textbf{76.51$\pm$11.87} & 74.55$\pm$13.55                \\  
\textbf{Pancreas}         & {85.00$\pm$5.95}                                                                           & {56.35$\pm$21.93} & \textcolor{red}{{75.78$\pm$8.68}}  & {81.9$\pm$6.88}   & {81.66$\pm$7.12}  & {81.46$\pm$9.11}  & \textbf{82.59$\pm$7.54}                 \\  
\textbf{Duodenum}         & {67.73$\pm$16.75}                                                                          & {38.39$\pm$22.04} & \textcolor{red}{{62.03$\pm$15.85}} & {63.96$\pm$16.44} & {66.66$\pm$16.18} & {66.55$\pm$16.02} & \textbf{68.23$\pm$15.04}                \\  
\textbf{Colon}                  & {87.26$\pm$8.25}                                                                           & {71.54$\pm$12.12} & \textcolor{red}{{78.77$\pm$10.61}}  & {83.77$\pm$8.45}  & {83.51$\pm$7.68}  & \textbf{85.61$\pm$7.19}  & 84.22$\pm$7.32                 \\  
\textbf{Intestine}              & {89.37$\pm$3.11}                                                                           & {72.44$\pm$8.26}  & \textcolor{red}{{72.8$\pm$8.62}}   & {86.38$\pm$3.77}  & {86.95$\pm$3.11}  & \textbf{87.36$\pm$3.55}  & 87.19$\pm$3.06                 \\  
\textbf{Adrenal}                & {72.98$\pm$8.09}                                                                           & {25.41$\pm$20.05} & \textcolor{red}{{60.55$\pm$11.13}}  & {68.26$\pm$7.83}  & {66.73$\pm$8.13}  & \textbf{70.08$\pm$8.67}  & 69.82$\pm$8.54                 \\  
\textbf{Rectum}                 & {82.32$\pm$5.26}                                                                           & {72.48$\pm$9.86}  & \textcolor{red}{{74.32$\pm$15.16}}  & {79.24$\pm$7.09}  & {79.26$\pm$8.57}  & \textbf{80.64$\pm$7.21}  & 79.99$\pm$6.82                 \\  
\textbf{Bladder}                & {92.11$\pm$9.75}                                                                           & \textbf{89.83$\pm$8.59}  & \textcolor{red}{{78.58$\pm$29.82}} & {87.54$\pm$17.11} & {88.18$\pm$15.91} & {87.6$\pm$15.88}  & 88.18$\pm$17.64                \\  
\textbf{Head of Femur (L)}       & {92.56$\pm$4.19}                                                                           & {84.32$\pm$4.98}  & \textcolor{red}{{88.24$\pm$2.81}}   & {91.71$\pm$4.44}  & {91.99$\pm$4.61}  & {91.74$\pm$4.36}  & \textbf{92.48$\pm$3.75}                 \\  
\textbf{Head of Femur (R)}       & {92.49$\pm$4.03}                                                                           & {89.12$\pm$2.69}  & \textcolor{red}{{89.04$\pm$2.42}}   & {92.04$\pm$3.22}  & {92.55$\pm$3.93}  & {92.0$\pm$3.58}   & \textbf{93.23$\pm$3.46}                 \\  
\hline
\textbf{Mean}                   & {87.41$\pm$4.57}                                                                           & {70.45$\pm$7.29}  & \textcolor{red}{{79.92$\pm$8.01}}  & {84.74$\pm$5.65}  & {85.1$\pm$5.09}   & {85.31$\pm$5.02}  & \textbf{85.82$\pm$4.89}                 \\ 
\hline
\end{tabular}
}
\caption{{Quantitative comparison between various efficient segmentation methods in term of $DSC~(\%)$. The teacher network is the well-trained nnUNetV2(3D). \textcolor{red}{We corrected the results or description by the red text.}}}
\label{tab:efficient_dsc}
\end{table}

\begin{table}[thp]
\centering
\normalsize
\scalebox{0.44}{
\begin{tabular}{l|c|cc|cc|cr}
\hline
\textbf{Method} & \textbf{nnUNetV2(3D)}& \textbf{ESPNet}           & \textbf{ESPNet+KD}        &\textbf{DMFNet}           & \textbf{DMFNet+KD}        & \textbf{LCOVNet}          & \textbf{LCOVNet+KD} \\
\hline
\textbf{Liver}                  &{3.17$\pm$0.51}                                                                            &{22.33$\pm$16.56} & {3.89$\pm$1.12}   & {3.79$\pm$0.96}   & \textbf{3.45$\pm$0.84}   & {15.25$\pm$52.03} & 5.52$\pm$6.11                  \\\textbf{Spleen}          & {2.12$\pm$0.47}                                                                            & {8.98$\pm$11.86}  & {14.35$\pm$37.18} & {4.11$\pm$5.55}   & {4.63$\pm$7.24}   & \textbf{3.64$\pm$3.95}   & 4.32$\pm$7.09                  \\\textbf{Kidney (L) } & {2.46$\pm$0.7}                                                                             & {9.94$\pm$15.51}  & {4.52$\pm$4.65}   & \textbf{2.79$\pm$0.52}   & {2.91$\pm$0.64}   & {2.81$\pm$0.74}   & 3.19$\pm$0.83                  \\
\textbf{Kidney (R)}        & {2.24$\pm$0.47}                                                                            & {5.06$\pm$6.34}   & {2.87$\pm$0.91}   & {2.64$\pm$0.49}   & {2.61$\pm$0.48}   & \textbf{2.56$\pm$0.45}   & 2.81$\pm$0.52                  \\ 
\textbf{Stomach}          & {9.47$\pm$7.61}                                                                            & {16.8$\pm$14.66}  & {19.0$\pm$24.07}  & {9.41$\pm$5.5}    & {10.83$\pm$9.36}  & \textbf{9.1$\pm$5.49}    & 15.97$\pm$32.96                \\ 
\textbf{Gallbladder}      & {6.04$\pm$5.63}                                                                            & {20.37$\pm$18.03} & {12.26$\pm$13.69} & {6.57$\pm$8.71}   & \textbf{6.21$\pm$6.91}   & {6.42$\pm$4.86}   & 11.86$\pm$26.18                \\ 
\textbf{Esophagus}        & {5.83$\pm$4.64}                                                                            & {15.49$\pm$14.52} & {10.5$\pm$10.36}  & {6.39$\pm$4.83}   & {6.17$\pm$4.37}   & \textbf{5.97$\pm$3.23}   & 7.09$\pm$8.06                  \\ 
\textbf{Pancreas}         & {6.87$\pm$7.86}                                                                            & {25.18$\pm$22.03} & {11.49$\pm$9.72}  & \textbf{8.02$\pm$8.38}   & {8.23$\pm$8.84}   & {9.91$\pm$13.95}  & 8.4$\pm$10.69                  \\ 
\textbf{Duodenum}         & {21.15$\pm$14.26}                                                                          & {41.61$\pm$19.34} & {32.64$\pm$30.05} & {23.28$\pm$14.73} & {20.32$\pm$13.11} & {20.65$\pm$13.35} & 19.4$\pm$11.39                 \\ 
\textbf{Colon}                  & {10.42$\pm$14.27}                                                                          & {73.68$\pm$83.07} & {20.68$\pm$13.62} & {11.18$\pm$11.71} & {11.11$\pm$12.13} & \textbf{9.55$\pm$11.72}  & 12.51$\pm$13.22                \\ 
\textbf{Intestine}              & {5.27$\pm$4.29}                                                                            & {17.47$\pm$8.02}  & {17.64$\pm$7.75}  & {6.85$\pm$4.34}   & {6.59$\pm$4.02}   & \textbf{6.04$\pm$3.6}    & 6.82$\pm$3.86                  \\ 
\textbf{Adrenal}                & {5.43$\pm$3.82}                                                                            & {32.45$\pm$21.58} & {10.45$\pm$11.4}  & {6.3$\pm$3.24}    & {6.75$\pm$4.41}   & \textbf{5.82$\pm$4.44}   & 6.06$\pm$4.07                  \\ 
\textbf{Rectum}                 & {12.39$\pm$8.12}                                                                           & {18.35$\pm$8.25}  & {18.87$\pm$19.96} & {11.41$\pm$5.51}  & {12.48$\pm$6.14}  & \textbf{10.16$\pm$4.55}  & 12.07$\pm$6.68                 \\ 
\textbf{Bladder}                & {4.17$\pm$3.6}                                                                             & \textbf{5.1$\pm$2.74}    & {20.03$\pm$50.47} & {5.93$\pm$6.02}   & {5.3$\pm$4.87}    & {6.89$\pm$6.67}   & 6.24$\pm$8.67                  \\ 
\textbf{Head of Femur (L)}       & {17.05$\pm$62.15}                                                                          & {12.69$\pm$4.91}  & {22.97$\pm$58.09} & \textbf{6.52$\pm$6.52}   & {6.58$\pm$6.61}   & {17.68$\pm$58.9}  & 17.09$\pm$55.62                \\ 
\textbf{Head of Femur (R)}       & {27.29$\pm$81.62}                                                                          & {9.41$\pm$3.49}   & {18.18$\pm$21.9}  & {6.14$\pm$4.33}   & \textbf{6.1$\pm$5.6}     & {30.83$\pm$93.82} & 6.5$\pm$5.78                   \\ 
\textbf{Mean}                   & {8.84$\pm$ 22.63}                                                                          & {20.93$\pm$18.13} & {15.02$\pm$16.3}  & {7.58$\pm$3.72}   & \textbf{7.52$\pm$3.59}   & {10.21$\pm$25.87} & 9.11$\pm$13.87                 \\
\hline
\end{tabular}

}
\caption{{Quantitative comparison between various efficient segmentation methods in term of $HD_{95}~(mm)$. The teacher network is the well trained nnUNetV2(3D).}}
\label{tab:efficient_hd}
\end{table}

\begin{table*}[thp]
\centering
\normalsize
\scalebox{0.47}{\begin{tabular}{l|c|c|c|c|c|c|c|c|c|c|c|c|c|c|c|c|c}
\hline
Method& Background & Liver& Spleen ~& ~ Kidney (L) ~& ~ Kidney (R)& ~ Stomach& ~ Gallbladder& ~Esophagus& ~ Pancreas& ~ Duodenum& Colon& Intestine& Adrenal& Rectum& Bladder& Head of Femur (L)& Head of Femur (R)\\
\hline
Ratio (\%)&0.23$\pm$0.03 &
1.79$\pm$0.3 &
2.82$\pm$0.51 &
1.89$\pm$0.51 &
1.95$\pm$0.42 &
1.97$\pm$0.34 &
7.32$\pm$3.75 &
9.99$\pm$3.9 &
4.57$\pm$2.11 &
5.9$\pm$3.23 &
1.8$\pm$0.36 &
1.75$\pm$0.33 &
19.76$\pm$4.34 &
2.35$\pm$1.19 &
1.58$\pm$1.19 &
1.23$\pm$0.25 &
1.08$\pm$0.28\\
\hline
\end{tabular}}
\caption{Labeling cost of scribble annotation compared with dense annotation. Here, we reported the percentage of labeled voxels between scribble and dense annotation.}
\label{tab:cost_analysis}
\end{table*}

\subsection{Abdominal organ segmentation with low computational cost and high speed}
Although large-scale deep models have achieved promising results for abdominal organ segmentation~\citep{isensee2021nnu,zhou2019prior,tang2021high}, these heavy models require various expensive computations and storage components and a long inference time~\citep{qin2021efficient}. In addition, the whole abdominal CT image has a very high resolution, which further increases the GPU memory budget and computational cost~\citep{qin2021efficient,tang2021high}. So, it is desirable to investigate the high-performance and low computational cost method for abdominal organ segmentation, and it is also suitable for clinical scenarios. This study investigates the efficient abdominal organ segmentation topic and compares several lightweight network-based and knowledge distillation methods on the WORD. Firstly, we compare three lightweight segmentation networks' performance in abdominal organ segmentation, ESPNet~\citep{espnet2018}, DMFNet~\citep{dmfnet2019} and LCOVNet~\citep{lcovnet2021}. ESPNet~\citep{espnet2018} proposed an efficient spatial pyramid block for high-speed brain tumour segmentation. DMFNet~\citep{dmfnet2019} combined point-wise~\citep{zhang2018shufflenet},
group-wise~\citep{chen2018multi} and atrous~\citep{chen2017deeplab} convolutions to reduce the computational cost and boost brain tumour segmentation performance. LCOVNet~\citep{lcovnet2021} proposed an attention based spatiotemporal separable convolution for efficient COVID-19  pneumonia lesion segmentation. Afterwards, we study the knowledge distillation strategy for the efficient high-resolution image segmentation~\citep{hinton2015distilling}. In general, knowledge distillation aims to transfer the knowledge of a heavy model (teacher) to a lightweight model (student) and encourages the student to achieve similar or comparable results to the teacher. Following the general knowledge distillation~\citep{hinton2015distilling,qin2021efficient}, we used a pre-trained nnUNetV2(3D) as the teacher model and employed the logit output of nnUNetV2(3D) to guide the student models (ESPNet, DMFNet and LCOVNet). Table~\ref{tab:efficient_dsc} and~\ref{tab:efficient_hd} list the quantitative results of different methods in terms of $DSC$ and $HD_{95}$. It can be observed that the knowledge distillation strategy can improve student models' performance. In Table~\ref{tab:para_ana}, we further analyze the model complexity of the teacher network and student networks in the same software and hardware environments~\footnote{https://github.com/sovrasov/flops-counter.pytorch}. {These results show that combining lightweight networks and knowledge distillation can achieve a better trade-off between performance and computational cost.} This study further indicates that exploring more power lightweight networks and knowledge distillation strategies is a potential solution for high-performance, fast-speed and low computational cost abdominal organ segmentation~\citep{feng2021resolution,qin2021efficient}.

\subsection{Abdominal organ segmentation with low annotation cost}
Recently, many annotation-efficient learning-based works have been employed to reduce medical image annotation cost~\citep{li2020transformation,luo2020semi,luo2021urpc,xia20203d}. However, most of them are semi-supervised learning-based methods, which still need to annotate part of the dataset carefully. Weakly supervised learning just requires a few sparse annotations to learn and achieve promising results~\citep{lin2016scribblesup,valvano2021learning}. Figure~\ref{fig:annotation_diff} shows an example of different types of medical image annotation. {Table~\ref{tab:cost_analysis} lists the percentage of labeled voxels of scribble annotation compared with dense annotation. It shows that sparse annotation can be used to produce coarse segmentation results with very few labeling cost.} In this work, we evaluate several weakly-supervised methods on the abdominal multi-organ segmentation task for the first time and further propose a new method to boost the results.

\begin{table}[ht]
\centering
\normalsize
\scalebox{0.48}{
\begin{tabular}{l|c|c|c|c|c|r}
\hline
\textbf{Method}                   & \textbf{pCE}                                        & \textbf{pCE+CRF Loss}                               & \textbf{pCE+EM}                                    & \textbf{pCE+IVM}          & \textbf{pCE+EM+IVM}                                 & \textbf{Mask} \\
\hline
\textbf{Liver}                  & {\textbf{93.86$\pm$0.88}}  & {73.28$\pm$3.51}                            & {92.62$\pm$1.67}                           & {88.46$\pm$2.48}  & {90.22$\pm$1.92}                            & 96.55$\pm$0.89                   \\ 
\textbf{Spleen }          & {89.43$\pm$4.27}                            & {81.71$\pm$8.16}                            & {87.25$\pm$5.98}                           & {88.67$\pm$6.69}  & {\textbf{91.42$\pm$3.4}}   & 95.26$\pm$2.84                   \\ 
\textbf{Kidney (L) } & {87.68$\pm$6.48}                            & {\textbf{92.46$\pm$4.58}}  & {88.68$\pm$3.36}                           & {92.02$\pm$2.89}  & {92.13$\pm$2.47}                            & 95.63$\pm$1.2                    \\ 
\textbf{Kidney (R)}        & {90.02$\pm$4.11}                            & {\textbf{92.84$\pm$4.17}}  & {88.49$\pm$3.78}                           & {90.59$\pm$2.55}  & {92.07$\pm$2.78}                            & 95.84$\pm$1.16                   \\ 
\textbf{Stomach}          & {87.09$\pm$4.24}                            & {86.64$\pm$4.3}                             & {\textbf{87.38$\pm$3.61}} & {86.98$\pm$4.44}  & {86.17$\pm$2.89}                            & 91.58$\pm$2.86                   \\ 
\textbf{Gallbladder}      & {62.13$\pm$18.78}                           & {63.51$\pm$20.52}                           & {65.21$\pm$18.94}                          & {65.09$\pm$16.67} & {\textbf{70.64$\pm$18.19}} & 82.83$\pm$11.8                   \\ 
\textbf{Esophagus}        & {34.99$\pm$10.7}                            & {55.53$\pm$13.77}                           & {41.2$\pm$13.38}                           & {54.22$\pm$13.09} & {\textbf{62.53$\pm$13.1}}  & 77.17$\pm$14.68                  \\ 
\textbf{Pancreas}         & {72.27$\pm$7.26}                            & {75.27$\pm$7.34}                            & {72.66$\pm$7.4}                            & {74.3$\pm$7.15}   & {\textbf{76.2$\pm$6.66}}   & 83.56$\pm$5.6                    \\ 
\textbf{Duodenum}         & {52.37$\pm$11.07}                           & {56.59$\pm$12.57}                           & {57.7$\pm$13.05}                           & {55.06$\pm$12.16} & {\textbf{58.47$\pm$13.15}} & 66.67$\pm$15.36                  \\ 
\textbf{Colon}                  & {72.65$\pm$10.04}                           & {66.95$\pm$10.68}                           & {76.03$\pm$9.9}                            & {74.21$\pm$9.84}  & {\textbf{78.66$\pm$9.38}}  & 83.57$\pm$8.69                   \\ 
\textbf{Intestine}              & {75.37$\pm$5.28}                            & {69.71$\pm$5.74}                            & {76.56$\pm$5.17}                           & {75.07$\pm$4.31}  & {\textbf{80.44$\pm$3.67}}  & 86.76$\pm$3.56                   \\ 
\textbf{Adrenal}                & {36.26$\pm$10.2}                            & {\textbf{46.09$\pm$10.27}} & {31.44$\pm$10.04}                          & {39.86$\pm$11.03} & {43.46$\pm$10.79}                           & 70.9$\pm$10.12                   \\ 
\textbf{Rectum}                 & {\textbf{70.77$\pm$10.61}} & {28.66$\pm$14.53}                           & {70.47$\pm$9.86}                           & {71.2$\pm$8.08}   & {69.62$\pm$9.55}                            & 82.16$\pm$6.73                   \\ 
\textbf{Bladder}                & {82.77$\pm$13.92}                           & {\textbf{87.07$\pm$16.42}} & {83.79$\pm$16.43}                          & {78.76$\pm$13.58} & {68.52$\pm$11.17}                           & 91.0$\pm$13.5                    \\ 
\textbf{Head of Femur (L)}       & {73.12$\pm$8.65}                            & {\textbf{86.5$\pm$4.33}}   & {80.39$\pm$7.58}                           & {82.41$\pm$5.24}  & {83.85$\pm$3.7}                             & 93.39$\pm$5.11                   \\ 
\textbf{Head of Femur (R)}       & {72.19$\pm$8.18}                            & {\textbf{87.73$\pm$4.14}}  & {82.6$\pm$6.74}                            & {82.92$\pm$5.34}  & {84.53$\pm$3.21}                            & 93.88$\pm$4.30                   \\ 
\hline
\textbf{Mean}                   & {72.06$\pm$19.78}                           & {71.91$\pm$20.53}                           & {73.90$\pm$19.57}                          & {74.99$\pm$17.02} & {\textbf{76.81$\pm$16.01}} & 86.67$\pm$4.81                   \\
\hline
\end{tabular}
}
\caption{Comparison between various weakly-supervised segmentation methods in the term of $DSC (\%)$, all methods based on the same backbone the ResUNet (2D) and same experiment settings.}
\label{tab:weak_dsc}
\end{table} 

\begin{table}
\centering
\normalsize
\scalebox{0.8}{\begin{tabular}{c|c|c|c}
\hline
    Network&Params (M)&MAC (G)&Inference Time (s)\\
    \hline
    nnUNetV2(3D)&31.18&580.77&0.47\\
    DMFNet&3.87&21.5&0.28\\
    ESPNet&4.45&458.78&0.29\\
    LCOVNet&0.82&100.21&0.21\\
\hline
\end{tabular}}
\caption{Complexity comparison between various networks. Params and MACs mean the model parameters and multiply-accumulate operations. The MACs and Inference Time were tested on an NVIDIA GTX1080TI GPU with the input size of 64$\times$160$\times$160.}
\label{tab:para_ana}
\end{table} 

\begin{table}[ht]
\centering
\normalsize
\scalebox{0.485}{
\begin{tabular}{l|c|cccc|r}
\hline
\textbf{Method}                   & \textbf{pCE}                                        & \textbf{pCE+CRF Loss}                               & \textbf{pCE+EM}                                    & \textbf{pCE+IVM}          & \textbf{pCE+EM+IVM}                                 & \textbf{Mask} \\
\hline
\textbf{Liver}                  & {\textbf{7.84$\pm$9.3}}   & {29.48$\pm$7.07}                            & {16.13$\pm$16.49}  & {17.85$\pm$18.14}                          & {9.47$\pm$2.44}                             & 4.64$\pm$7.37                    \\   
\textbf{Spleen }          & {9.35$\pm$9.29}                            & {25.2$\pm$40.37}                            & {20.78$\pm$39.72}  & {10.26$\pm$11.13}                          & {\textbf{8.33$\pm$9.55}}   & 8.7$\pm$30.11                    \\   
\textbf{Kidney (L) } & {39.23$\pm$110.58}                         & {13.79$\pm$27.9}                            & {19.7$\pm$29.74}   & {\textbf{7.37$\pm$14.57}} & {7.61$\pm$17.25}                            & 5.4$\pm$15.85                    \\   
\textbf{Kidney (R)}        & {31.68$\pm$45.25}                          & {9.41$\pm$20.34}                            & {58.63$\pm$151.23} & {12.06$\pm$24.16}                          & {\textbf{8.39$\pm$22.65}}  & 2.47$\pm$0.97                    \\   
\textbf{Stomach}          & {13.43$\pm$8.69}                           & {14.43$\pm$10.76}                           & {19.83$\pm$21.43}  & {12.92$\pm$7.78}                           & {\textbf{12.60$\pm$7.05}}  & 9.98$\pm$6.62                    \\   
\textbf{Gallbladder}      & {31.28$\pm$27.84}                          & {\textbf{11.29$\pm$11.26}} & {47.52$\pm$128.15} & {32.57$\pm$25.49}                          & {15.04$\pm$12.79}                           & 9.48$\pm$12.97                   \\   
\textbf{Esophagus}        & {24.9$\pm$10.02}                           & {12.69$\pm$9.13}                            & {20.61$\pm$9.16}   & {15.13$\pm$9.09}                           & {\textbf{12.51$\pm$9.23}}  & 6.7$\pm$7.6                      \\   
\textbf{Pancreas}         & {11.94$\pm$10.91}                          & {\textbf{10.26$\pm$8.96}}  & {13.43$\pm$9.3}    & {12.58$\pm$10.64}                          & {10.5$\pm$8.13}                             & 7.82$\pm$7.15                    \\   
\textbf{Duodenum}         & {36.36$\pm$17.15}                          & {22.34$\pm$12.66}                           & {22.64$\pm$12.18}  & {25.33$\pm$17.05}                          & {\textbf{22.25$\pm$11.43}} & 21.79$\pm$12.83                  \\   
\textbf{Colon}                  & {27.03$\pm$14.03}                          & {\textbf{18.04$\pm$12.63}} & {25.11$\pm$14.69}  & {23.85$\pm$14.02}                         & {18.55$\pm$13.77}                           & 17.41$\pm$15.22                  \\   
\textbf{Intestine}              & {18.47$\pm$8.77}                           & {17.6$\pm$5.58}                             & {19.28$\pm$11.33}  & {19.77$\pm$7.64}                           & {\textbf{12.02$\pm$5.56}}  & 9.54$\pm$7.2                     \\  
\textbf{Adrenal}                & {23.6$\pm$10.32}                           & {\textbf{13.02$\pm$6.93}}  & {24.93$\pm$10.32}  & {20.86$\pm$9.45}                           & {14.64$\pm$7.15}                            & 6.67$\pm$4.59                    \\   
\textbf{Rectum}                 & {\textbf{11.99$\pm$5.59}} & {22.37$\pm$10.43}                           & {21.2$\pm$12.32}   & {14.42$\pm$9.07}                           & {12.59$\pm$6.08}                            & 10.62$\pm$6.52                   \\   
\textbf{Bladder}                & {21.94$\pm$26.26}                          & {\textbf{6.71$\pm$5.86}}   & {17.13$\pm$24.9}   & {13.12$\pm$8.59}                           & {17.58$\pm$9.99}                            & 5.02$\pm$6.17                    \\   
\textbf{Head of Femur (L)}       & {72.85$\pm$99.82}                          & {20.97$\pm$58.39}                           & {34.85$\pm$95.43}  & {33.34$\pm$95.46}                          & {\textbf{20.05$\pm$60.58}} & 6.56$\pm$8.3                     \\   
\textbf{Head of Femur (R)}       & {51.87$\pm$93.12}                          & {20.58$\pm$59.5}                            & {33.44$\pm$94.53}  & {33.08$\pm$94.72}                          & {\textbf{19.2$\pm$58.89}}  & 5.98$\pm$7.2                     \\ 
\hline
\textbf{Mean}                   & {27.11$\pm$34.90}                          & {16.76$\pm$17.41}                           & {25.95$\pm$45.48}  & {19.03$\pm$27.55}                          & {\textbf{13.83$\pm$17.03}} & 8.6$\pm$6.47                     \\ 
\hline
\end{tabular}

}
\caption{Comparison between various weakly-supervised segmentation methods in the term of $HD_{95}~(mm)$, all methods based on the same backbone ( ResUNet (2D)) and same experiment settings.}
\label{tab:weak_hd95}
\end{table} 

\subsubsection{Learning from scribbles}
To learn from scribble annotations, the general method is using the partially Cross-Entropy (pCE) loss to train deep networks, where just labelled pixels are considered to calculate the gradient and the other pixels are ignored~\citep{tang2018normalized}. However, due to the sparse supervision, the pCE loss can not achieve promising results. To solve this dilemma,~\cite{tang2018regularized} proposed to integrate the pCE loss and MRF/CRF regularization terms to train deep networks with scribble annotations. After that, most of the recent weakly-supervised methods trained deep networks by using the following joint objective function~\citep{valvano2021learning,zhang2020weakly}:
\begin{equation}
    \mathcal{L}_{total} = \mathcal{L}_{pCE}+\lambda_{1}\mathcal{L}_{CRF}+\lambda_{2}\mathcal{L}_{other}
\end{equation}where $\mathcal{L}_{other}$ means other loss functions presented in these works. $\lambda_{1}$ and $\lambda_{2}$ denote the weight factor of these loss functions. These methods have achieved encouraging results in natural image segmentation\citep{tang2018normalized,tang2018regularized} and salience object detection~\citep{zhang2020weakly,yu2021structure}, etc. But for abdominal multi-organ segmentation, learning from scribbles is also a very challenging task. Different from the above, we propose a new regularization term to train deep networks for weakly supervised abdominal multi-organ segmentation.

\begin{table}
\centering
\normalsize
\scalebox{0.7}{\begin{tabular}{l|c|r}
\hline
    Method (with \textit{n} $\times$ \textit{n} dilation kernel) & $DSC~(\%)$ &$HD_{95}~(mm)$\\
    \hline
    pCE~(None)&72.06$\pm$19.78&27.11$\pm$34.90\\
    pCE+EM+IVM~(None)&76.81$\pm$16.01&13.83$\pm$17.03\\
    \hline
    pCE~(3 $\times$ 3)&74.23$\pm$18.49&23.93$\pm$18.65\\
    pCE+EM+IVM~(3 $\times$ 3)&78.32$\pm$14.57&12.57$\pm$16.09\\
    \hline
    pCE~(5 $\times$ 5)&75.93$\pm$19.35&19.48$\pm$21.03\\
    pCE+EM+IVM~(5 $\times$ 5)&79.68$\pm$15.56&12.74$\pm$13.37\\
    \hline
    pCE~(7 $\times$ 7)&77.86$\pm$16.48&15.62$\pm$18.19\\
    pCE+EM+IVM~(7 $\times$ 7)&80.71$\pm$12.26&12.96$\pm$12.83\\
    
\hline
\end{tabular}}

\caption{Sensitivities to scribble thickness evaluated on the WORD dataset testing set. The dilated scribbles are simulated from the origin scribbles, expanding their thickness by dilating with different kernels. Here, we reported the mean results of 16 organs in terms of $DSC$ and $HD_{95}$.}
\label{tab:scribble_ratio}
\end{table} 

\subsubsection{Entropy minimization}
Recently, entropy minimization has been widely used in semi-supervised learning to utilize the unlabeled data~\citep{grandvalet2005semi,hang2020local,vu2019advent}. It encourages the model to produce high confidence prediction by minimizing the following object function:
\begin{equation}
    \mathcal{L}_{ent} = \sum_{c} \sum_{i} -p_c^i \cdot \log{p_c^i}
\end{equation}where $p_c^i$ means the probability value of the pixel $i$ belonging to the $c$ class. In this work, we further use entropy minimization to regularize the deep network for learning from scribble annotations. Our intuition is that the entropy minimization is more like pixel-wise contrastive learning to encourage the model to learn from unlabeled pixels by min-max the intra/inter-class discrepancy. As the softmax prediction has maximized the difference of inter-class and the entropy minimization term enforces the intra-class prediction to be confident. 

\subsubsection{Intra-class intensity variance minimization}
Although the entropy minimization loss has regularized the deep network at the output level, it does not consider the image-level information. We hypothesize that the intensity information could bring more useful information and further boost model performance. Here, we attempt to reformate an unsupervised regularization term to consider both prediction and intensity simultaneously. Inspired by the clustering learning~\citep{jain1988algorithms} and active contour model~\citep{chan2001active}, we propose to regularize the deep network by minimizing the intra-class intensity variance, where the mathematical formulation is defined as:
\begin{equation}
    \mathcal{L}_{ivm} = \int(p_c^i \cdot I^i - u_c)^2 didc
\end{equation}where
\begin{equation}
    u_c =  \frac{\int (I^i \cdot p_c^i)di}{\int p_c^i di}
\end{equation}where $I^i$ denotes the intensity value of the input image at pixel $i$. $c$ is the class number. Based on the above descriptions, the $\mathcal{L}_{ivm}$ can be converted to the intra-class intensity standard deviation minimization term (\textit{std}).


\subsubsection{The overall objective function}
In this work, we employ a joint objective function to train model from the scribble annotations, which consists of three terms: partially cross-entropy loss, entropy minimization loss and intensity variance minimization loss and takes the following combination:
\begin{equation}
    \mathcal{L}_{total} = \mathcal{L}_{pCE}+\lambda_{ent}\mathcal{L}_{ent}+\lambda_{ivm}\mathcal{L}_{ivm}
\end{equation}where $\lambda_{ent}$ and $\lambda_{ivm}$ represent the importance of $\mathcal{L}_{ent}$ and $\mathcal{L}_{ivm}$ respectively and both are set to 0.1 in this work.

\subsubsection{Experiments and results}
\textbf{\textit{Experiments settings: }}To evaluate the proposed method, we further provide scribble annotations for the WORD. We generate scribbles for all training volumes in the axial view. Note that the scribble annotations are very sparse in both intra-/inter-slices, which means that not all slices have the scribbles but each organ is annotated at least once in a volume. Due to the scribble annotations based abdominal multi-organ segmentation is not a hot research topic, there is no existing work or openly available codebase. We first build a benchmark for this task and then compare a widely-used method, CRFLoss~\citep{tang2018regularized} on the WORD. We use the ResUNet (2D)~\citep{diakogiannis2020resunet} as our backbone and employ the nnUNet~\citep{isensee2021nnu} pipeline to train and test all methods. All implementations and scribbles are released.

\textbf{\textit{Results: }}The quantitative comparisons between our proposed method and the others are presented in Table~\ref{tab:weak_dsc} and Table~\ref{tab:weak_hd95}. The first interesting observation is that the widely-used CRF Loss~\citep{tang2018regularized} achieves a worst performance than all other methods. The reason may be that the CRF Loss~\citep{tang2018regularized} is specifically designed for natural image segmentation tasks and is not suitable for handling CT images with low contrast and non-enhancement. Then, we found that the network can leverage the scribble annotation more efficiently by encouraging to produce more confident predictions. Moreover, compared with the entropy minimization term, our proposed intra-class intensity variance minimization achieves better results, the mean $DSC$ of 73.90\% $vs$ 74.99\%. In addition, by combining the entropy minimization and intra-class intensity variance minimization, the model achieves the best performance of the others and improves the mean $DSC$ from 72.06\% to 76.81\%. These results demonstrate that also most weakly supervised methods achieve better results than those using partially cross-entropy loss, except for the CRF Loss. It is noteworthy to mention that scribble annotations save more than 96\% of labeling costs than dense annotations. In addition, we find large size organs weakly supervised segmentation results are very close to fully supervised, especially in the femur's liver, spleen, kidney, stomach, and head. However, the small size organs still can not be segmented well, it also points out the research direction. {Moreover, we further investigate the network performance when increasing the scribble thickness in Table~\ref{tab:scribble_ratio}, where we increase the thickness by dilating original scribbles with 3 $\times$ 3, 5 $\times$ 5, and 7 $\times$ 7 kernels in the axial view. It shows that the proposed method has a higher performance when increasing the scribble thickness, suggesting that the proposed can benefit from the increased scribble thickness. The above results show that weakly-supervised learning may further reduce the labeling cost with further research.}

\begin{table}[thp]
\centering
\normalsize
\scalebox{0.51}{\begin{tabular}{l|c|c|c|c|c|c|c|c}
\hline
Method&\textit{Open}&\textit{Tr}/\textit{Ts}&~Gallbladder~& ~Esophagus& ~ Pancreas& ~ Duodenum&~Adrenal& Rectum\\
\hline
\citet{oliveira2018novel}&Yes&20/10&51.8&-&57.2&-&-&-\\
\citet{wang2019abdominal}&No&236*&90.5&-&87.8&75.4&-&-\\
\citet{tang2021high}&Yes&30/20&82.6&78.8&76.1&-&73.6&-\\
\citet{liang2021incorporating}&Yes&90**&78&74&81&71&-&-\\
\citet{chen2021deep}&No&150/20&87&76&84&77&-&80\\
\hline
Ours&Yes&120/50&83.2&78.5&85.0&68.3&72.4&82.4\\
\hline
\end{tabular}}
\caption{Segmentation results of existing methods for small abdominal organs segmentation in CT scan (gallbladder, esophagus, pancreas, duodenum, adrenal, rectum). * and ** mean four-fold and nine-fold cross-validation, respectively. \textit{Tr} and \textit{Ts} represent the total cases of the training set and testing set, respectively. \textit{Open} denotes the data is open available.}
\label{tab:liture_review}
\end{table}

\section{Discussion and Conclusion}
In this work, we collect and build a large-scale whole abdominal CT multi-organ segmentation dataset containing 150 CT volumes and 16 organ annotations. Although, many abdominal organ segmentation datasets and benchmarks have been established, like AbdomenCT-1K~\citep{ma2021abdomenct}, BTCV~\citep{landman2017multi},  TCIA~\citep{roth2015deeporgan}, LiTS~\citep{bilic2019liver}, CT-ORG~\citep{rister2020ct}, KiTS~\citep{heller2020international}, etc, our WORD dataset cover the whole abdominal region and also annotate more organs. Then we annotate 20 scans from the open available ~\citep{bilic2019liver} for clinical applicable and generalizable evaluation. Here, we investigate several hot topics based on the WORD dataset and point out some unsolved or challenging problems.

\subsection{Clinical applicable investigation }{We investigate several SOTA methods on the WORD dataset and find that all methods can achieve encouraging results. Then, we comprehensively study the clinical acceptance of the deep network. Figure~\ref{fig:revision_degree} shows three junior clinical oncologists revise the results. For large-scale organs, such as the liver, spleen, kidney, stomach, bladder, and head of the femur, the deep network can perform very closely to junior oncologists, which means the model prediction can be clinically acceptable after minor revision. However, there are huge performance gaps between junior oncologists and the deep network on small organs, such as the gallbladder, esophagus, pancreas, duodenum, adrenal, and rectum, suggesting that directly applying the model predictions to the clinical application is tough without oncologists revision. Moreover, we further investigate the segmentation results of existing methods for these challenging and small organs. Quantitative results are listed in Table~\ref{tab:liture_review}. It is worth pointing out that the comparisons are unfair, as the dataset and experimental settings for each method are different. It can be found that the results in WORD dataset are more comprehensive and competitive than those in recent works~\citep{oliveira2018novel,wang2019abdominal,tang2021high,liang2021incorporating,chen2021deep}, but it's still not good enough for clinical application. So, we think the abdominal multi-organ segmentation task is not a well-solved problem. And the WORD dataset not only can provide a fair benchmark for performance comparison but also help researchers focus on handling these challenging organ segmentation to improve clinical practice performance.}

\subsection{Model generalization }Recently, domain adaptation and generalization have been scorching topics in the natural/medical image segmentation fields~\citep{dou2018unsupervised,vu2019advent}. But for the abdominal multi-organ segmentation task, there are very few studies~\citep{dou2020unpaired} focused on investigating the generalization problem. This is mainly due to lacking open available multi-sources and large-scale datasets/benchmarks. In this work, we investigate the domain gaps between our build WORD dataset and open-source datasets BTCV~\citep{landman2017multi} and TCIA~\citep{roth2015deeporgan} and find that there are significant domain gaps across different source datasets. Furthermore, we annotated 20 volumes from LiTS~\citep{bilic2019liver} as an external evaluation set to validate networks' generalizability and found that the domain gap between LiTS and WORD dataset is not significant. It is desirable to train models with good generalization and high-performance to boost deep learning-based clinical application. So, we build a benchmark for robust and generalizable abdominal multi-organ segmentation research.

\subsection{Annotation-efficient segmentation }Developing an encouraging performance segmentation model always requires many high-quality annotations, but labelling the abdominal multi-organ is very expensive and time-consuming, each volume around takes 1.2-2.6 hours. To reduce the labeling cost, annotation-efficient learning has attracted many researchers' attention, such as semi-supervised learning~\citep{luo2020semi,luo2021urpc,ssl4mis2020,media2022urpc,you2021momentum,you2022class,you2020unsupervised,you2022simcvd,you2022bootstrapping,you2022incremental} and weakly supervised learning~\citep{valvano2021learning,luo2022scribble}. In this work, we propose to learn from the scribble annotation by minimizing the entropy minimization and intra-class intensity variance minimization. Although our proposed method improves the baseline by a large margin, there is also a considerable performance gap compared with dense annotations. In this work, we want to do some attempts to inspire annotation-efficient research in the future.

\par In conclusion, we introduced a new carefully annotated whole abdominal organ CT dataset. Meanwhile, we investigate several existing SOTA methods and perform user study on this dataset, and further point out some unsolved problems and potential directions in both technique and clinical views. In the future, we will still work on extending the WORD dataset to be more extensive and more diverse.



\section{Acknowledgment}
This work was supported by the National Natural Science Foundation of China [81771921 , 61901084 ], the National Key Research and Development Program [2020YFB1711503] and also by key research and development project of Sichuan province, China [20ZDYF2817]. We would like to thank Mr. Zhiqiang Hu and Guofeng Lv from the SenseTime Research for constructive discussions and suggestions and also thank M.D. J. Xiao and W. Liao and their team members for data collection, annotation, checking and user study. We also would like to thank the Shanghai AI Lab and Shanghai SenseTime Research for their high-performance computation support.

\bibliographystyle{model2-names.bst}\biboptions{authoryear}
\bibliography{refs}



\end{document}